\documentclass[prd,twocolumn,aps,amsmath,nofootinbib,superscriptaddress]{revtex4}

\usepackage{graphicx}
\usepackage{amsfonts}
\usepackage[usenames]{color}
\usepackage{amsmath}

\newcommand{\mL}{{\mathcal L}}

\newcommand{\mO}{{\mathcal O}}

\newcommand{\bea}{\begin{eqnarray}}
\newcommand{\eea}{\end{eqnarray}}
\newcommand{\nn}{\nonumber}
\newcommand{\vev}[1]{ \left\langle {#1} \right\rangle }

\newcommand{\comment}[1]{}

\def\R{\mathbb{R}}

\def\SO{\mathop{\rm SO}}

\def\SU{\mathop{\rm SU}}
\def\U{\mathop{\rm U}}

\def\simgt{\mathrel{\lower2.5pt\vbox{\lineskip=0pt\baselineskip=0pt
           \hbox{$>$}\hbox{$\sim$}}}}
\def\simlt{\mathrel{\lower2.5pt\vbox{\lineskip=0pt\baselineskip=0pt
           \hbox{$<$}\hbox{$\sim$}}}}

\begin{document}


\preprint{IPMU08, UT-08-29}

\title{Neutrino mixing and mass hierarchy in Gaussian landscapes}

\author{Lawrence J.~Hall}
\affiliation{Department of Physics and Lawrence Berkeley National 
Laboratory,University of California, Berkeley, CA 94720, USA}

\author{Michael P.~Salem}
\affiliation{Institute of Cosmology, Department of Physics and Astronomy, \\
Tufts University, Medford, MA 02155, USA}

\author{Taizan Watari}
\affiliation{Department of Physics, University of Tokyo, Tokyo, 
113-0033, Japan} 
\affiliation{Institute for the Physics and Mathematics of the Universe 
(IPMU), University of Tokyo, Kashiwa-no-ha 5-1-5, 277-8592, Japan} 

\begin{abstract}
The flavor structure of the Standard Model may arise from random selection
on a landscape.  In a class of simple models, called ``Gaussian landscapes,''  
Yukawa couplings derive from overlap integrals of Gaussian zero-mode 
wavefunctions on an extra-dimensional space.  Statistics of vacua are 
generated by scanning the peak positions of these wavefunctions, giving 
probability distributions for all flavor observables.  Gaussian landscapes 
can account for all of the major features of flavor, including both the small 
electroweak mixing in the quark sector and the large mixing observed in the 
lepton sector.  We find that large lepton mixing stems directly from lepton 
doublets having broad wavefunctions on the internal manifold.  Assuming 
the seesaw mechanism, we find the mass hierarchy among neutrinos is sensitive 
to the number of right-handed neutrinos, and can provide a good fit to 
neutrino oscillation measurements.
\end{abstract}

\maketitle


\section{Introduction}

The Standard Model of particle physics, taken to include neutrino masses, 
is described by a quantum field theory with about two dozen input 
parameters.  The flavor observables---quark and lepton masses, mixings and 
CP phases---constitute a large fraction of these inputs, reflecting the 
fact that we have not found a simple principle (like gauge coupling 
unification) to relate all of them.  It is possible that there is no deep 
meaning behind the precise values of many of the flavor parameters that we 
observe.  Furthermore, the set of Yukawa couplings in our vacuum may be one 
among many possibilities that the fundamental (microscopic) theory admits.  
If this is the case, then one can understand the measured values of flavor 
observables only in terms of statistical distributions; still the large 
number of flavor observables gives hope for a discerning statistical 
analysis.  These ideas have been pursued in Refs.~\cite{DDR,HMW,HSW1,HSW2}.  
Note that in string theory, flux compactification succeeds in stabilizing 
some moduli, the statistics of fluxes in the internal space generates 
statistics of moduli values, and these may in turn generate statistics of 
Yukawa couplings~\cite{flux}.  Thus, the current understanding of string 
theory supports a statistical picture of flavor.

The Yukawa couplings of the Standard Model cannot be completely random
numbers.  The eigenvalues of Yukawa matrices are hierarchically separated, at 
least in the quark and charged lepton sectors, and the charged electroweak 
current pairs each left-handed up-type quark almost uniquely to a left-handed 
down-type quark; furthermore this pairing combines the heaviest up-type 
quark to the heaviest down-type quark, and similarly for the middle and 
lightest quarks.  These patterns are respectively referred to as hierarchy, 
pairing structure, and generation structure, and a statistical theory of 
flavor observables should explain each of them. 

An intuitive explanation of these patterns is available in string theory
compactification~\cite{HSW2}.  Fields in the low-energy effective theory 
(e.g.~the Standard Model) are Kaluza--Klein zero modes on some internal
manifold, and Yukawa couplings are calculated by overlap integration of 
the zero-mode wavefunctions of the three fields relevant to any coupling. 
Overlap integration of localized wavefunctions generates hierarchy~\cite{AS}.  
Localized wavefunctions of the quark doublets and the Higgs boson introduce 
correlation between the up-type and down-type Yukawa matrices, giving rise
to pairing structure and generation structure.   Meanwhile, localized 
zero-mode wavefunctions are fairly easy to obtain in torus-fibered 
compactifications with small torus fiber, see e.g.~Refs.~\cite{Witten}. 
These basic features of quark flavor are nicely explained in Gaussian 
landscapes~\cite{HSW1,HSW2}, which are proposed as toy models of the 
landscape, capturing the essential features of torus-fibered 
compactification of Heterotic string theory and its dual descriptions.

The observation of large mixing angle neutrino oscillations has been 
welcomed with surprise, as it reveals that the matter fields of the 
Standard Model are not simply three copies of a spinor representation of 
$\SO(10)$.  The lack of pairing structure in the lepton sector presumably 
indicates that flavor structure is quite different between quark and lepton 
sectors.  {\it How} do the quark and lepton sectors come to have different 
flavor structure, and {\it why}?  These questions point to a deeper, more 
fundamental understanding of the microscopic origin of flavor.  Such an 
understanding might influence the way we study leptogenesis, or how we make 
predictions for the flavor observables (such as $\theta_{13}^{\rm PMNS}$ 
and CP-violating phases) to be measured in future neutrino experiments. 

Conventional theories of flavor assume that mass hierarchies and mixing 
angles are determined by approximate flavor symmetries.  With the discovery 
of large neutrino mixing angles, and at most a modest neutrino mass 
hierarchy, Ref.~\cite{HMW} proposed a statistical ``neutrino anarchy" in the 
lepton sector, with lepton doublets not determined by any symmetry.  In 
the Gaussian landscape~\cite{HSW1,HSW2}, a microscopic origin for quark mass 
hierarchies and mixing angles emerges from the localization of zero modes 
on an internal manifold.  It was a natural guess in Ref.~\cite{HSW2} 
that large lepton mixing angles would follow from lepton doublets having 
less localized zero-mode wavefunctions than quark doublets,  allowing a 
description of ``{\em how}" the quark and lepton sectors differ.

We here conclude that this guess is indeed correct.  The present authors 
claimed in Refs.~\cite{HSW1,HSW2} that large mixing in the lepton sector 
required both non-localized wavefunctions of lepton doublets and large 
complex phases.  The latter were seen as necessary to prevent a tidy 
cancellation of terms in the PMNS matrix.  However, we have since found 
that the above cancellation proceeds as an artifact of ignoring kinetic 
mixing of fermions in Refs.~\cite{HSW1,HSW2}.  With kinetic mixings taken 
into account in Gaussian landscapes, we find that complex phases are not 
necessary for large lepton mixing.

We also find that the (statistical distributions of) low-energy neutrino 
masses, generated by the see-saw mechanism in Gaussian landscapes, are 
sensitive to the number of right-handed neutrinos.  When there is a large 
number of right-handed neutrinos, their mass eigenvalues become more densely 
packed, and the hierarchy between the $\Delta m^2$ of solar and 
that of atmospheric neutrino oscillations need not be large.  The existence 
of many right-handed neutrinos is a natural consequence of string theory 
compactification, because right-handed neutrinos are just 
$\SU(5)_{\rm GUT}$-singlet moduli fields, and there are often many moduli.  
Note that a dense concentration of right-handed neutrino masses also 
implies that more than one neutrino might contribute to thermal 
leptogenesis. 

The remainder of this paper is organized as follows.  The Gaussian landscape
is set up in Section~\ref{sec:background}, where we extend the construction 
of Refs.~\cite{HSW1,HSW2} to account for overlap integrals that may lead to 
non-canonical kinetic terms in the low-energy theory.  Our main results 
appear in Section~\ref{sec:numerical}, where we describe a numerical 
simulation of a Gaussian landscape and comment on its (in)sensitivity to
model assumptions.  In Section~\ref{sec:AFSapprox} we briefly outline how
certain analytical approximation methods, developed in Ref.~\cite{HSW2} to
provide insight into a Gaussian landscapes, can be extended to our revised
construction.  Concluding remarks are given in Section~\ref{sec:conclusions}.


\section{Gaussian Landscapes}
\label{sec:background}

\subsection{Kaluza--Klein Reduction}
\label{ssec:KKreduction}

Consider a low-energy effective theory (e.g.~the Standard Model) obtained 
by compactifying a supersymmetric Yang--Mills theory of group 
$G \supset \SU(3)_C \times \SU(2)_L \times \U(1)_Y \equiv G_{\rm SM}$ on a
higher-dimensional spacetime.  Let the internal manifold have $D$ dimensions, 
and denote it by $X$.  We assume a gauge-field background exists on $X$ such 
that the unbroken gauge symmetry is reduced from $G$ to $G_{\rm SM}$.  
Fermions in the low-energy effective theory may be identified with the 
Kaluza--Klein zero modes of gauginos. The irreducible decomposition of 
$\mathfrak{g}$ under $G_{\rm SM}$ may contain 
$({\bf 3},{\bf 2})^{+1/6}\!+(\bar{\bf 3},{\bf 1})^{-2/3}\!
+(\bar{\bf 3},{\bf 1})^{+1/3}\!+({\bf 1},{\bf 2})^{-1/2}\! 
+({\bf 1},{\bf 1})^{+1}$ of $G_{\rm SM}$, i.e.~a set of representations 
corresponding to a ``generation'' of Standard Model fermions.  In each 
irreducible component, the higher-dimensional gaugino $\Psi(x,y)$ has a 
Kaluza--Klein decomposition 
\bea
\Psi^a (x,y) = \sum_{i} \psi^{a}_i(x)\, \varphi^{a}_{i}(y)
+ \sum_{I} \psi^{a}_{I}(x)\, \varphi^{a}_{I}(y) \,,
\eea
where $a$ labels the irreducible representations listed above 
($a\!=\!q$, $\bar{u}$, $\bar{d}$, $\ell$, or $\bar{e}$), $i$ runs 
over the Kaluza--Klein zero modes, and $I$ runs over the massive modes. 
The wavefunctions $\varphi(y)$ denote mode functions of the Kaluza--Klein 
decomposition, while the $\psi(x)$ correspond to fields in the 3+1 
dimensional effective theory below the Kaluza--Klein scale ($x$ denotes 
coordinates in 3+1 dimensional Minkowski space, $y$ denotes those of the 
internal space $X$).  Note that the $\psi(x)$ are spinors of $\SO(3,1)$,
while the $\varphi(y)$ are spinors of $\SO(D)$, but spinor indices are 
suppressed.

If the topology of $X$ and the gauge-field background on $X$ are chosen 
appropriately, then there are three Kaluza--Klein zero modes in each 
irreducible representation, just like in the Standard Model.  In what 
follows, we only consider vacua with such properties.  If the irreducible 
decomposition of $\mathfrak{g}$ contains a singlet of $G_{\rm SM}$, then 
zero modes of this component may be identified with right-handed neutrinos. 
We extend the range of the label $a$ to include these right-handed 
neutrinos.  Because we do not require an unbroken $\SO(10)$ 
symmetry in the low-energy effective theory, the number of right-handed 
neutrinos is not necessarily three.  In fact, the right-handed neutrinos
are supersymmetric partners of gauge-field moduli in compactifications 
with supersymmetry, in which case it is quite likely that there are many 
of them.

The kinetic terms of the three fermions in each representation arise from 
dimensional reduction of the gaugino kinetic term in higher dimensions:
\bea
M_*^D\!\int_X d^D y\sqrt{g}\; \overline{\Psi}^a 
i\,\gamma^\mu\partial_\mu\Psi^a
\longrightarrow \sum_{i,j} K^a_{ij}  
\bar{\psi}^{a}_{i} i\,\gamma^\mu\partial_\mu\psi^{a}_{j}\,,
\eea
where
\bea
K^a_{ij} = M_*^D\!\int_X d^D y\sqrt{g}\; 
\overline{\varphi}^{a}_{i}(y)\,\varphi^{a}_{j}(y) \,.
\label{eq:gen-K}
\eea 
Here $M_*$ is the cutoff of the 4+D dimensional super Yang--Mills theory, 
and again $D$ is the number of extra dimensions. 

Meanwhile, the Higgs doublet of the Standard Model may be identified with 
the Kaluza--Klein zero mode of a gauge field $A_m(x,y)$ of $\mathfrak{g}$ 
on $X$.  The Kaluza--Klein decomposition of $A_m(x,y)$ may take the form 
\bea
A_m(x,y)= h(x)\, \varphi^h_m(y) + \sum_I h_I(x)\, \varphi_{Im}^h(y) 
+ {\rm h.c.}
\eea
Here $h(x)$ is a complex scalar field of the 3+1 dimensional effective 
theory and is in the ({\bf 1}, {\bf 2})$^{+1/2}$ representation of 
$G_{SM}$.  The zero-mode wavefunction of this scalar is $\varphi^h_m(y)$.
The Hermitian conjugate part contains $h^*(x)\,\varphi^h_m(y)^*$.
The Yukawa couplings of quarks and leptons then arise from dimensional 
reduction of the super Yang--Mills interactions on $X$:
\bea
&& \!\!\!\!
g_*M_*^D\! \int_X d^Dy\sqrt{g}\; \overline{\Psi}^{a}\Gamma^m A_m\Psi^b \nn\\
&& \qquad\qquad\qquad \longrightarrow \, 
\sum_{ij} \lambda^{ab}_{ij}\,\psi^a_i(x)\,h(x)\,\psi^b_j(x)\,, \qquad
\label{eq:setupYuk}
\eea
where 
\bea
\lambda^{ab}_{ij} = g_* M_*^D\! \int_X d^Dy\sqrt{g} \; 
   \varphi^a_i(y)\,\Gamma^m\varphi^h_m(y)\,\varphi^b_j(y) \,,   
\label{eq:gen-Yukawa}
\eea
for the up-type and neutrino Yukawa couplings, while replacing
$h(x)$ with $h^*(x)$ in Eq.~(\ref{eq:setupYuk}) and $\varphi^h_m(y)$ with 
$\varphi^h_m(y)^*$ in Eq.~(\ref{eq:gen-Yukawa}) for the down-type and 
charged-lepton Yukawa couplings.  Here $g_*$ is a 
dimensionless coupling constant of the gauge interaction of $G$ on the 
($D$+4) dimensional spacetime.  This is essentially how the Standard--Model 
Lagrangian is obtained in the compactification of Heterotic / Type I string 
theories.  Furthermore, some vacua of Type IIA / M-theory / Type IIB / 
F-theory compactifications are also approximated (albeit poorly) by this 
description, because of string duality.

The Kaluza--Klein zero modes of any irreducible component of the gaugino 
form a vector space, because the massless Dirac equation on a $D$ 
dimensional manifold $X$ is linear in $\varphi(y)$.  One is free to choose 
any basis for the vector space of these zero modes.  The kinetic mixing 
coefficients $K^a_{ij}$ and the Yukawa couplings $\lambda^{ab}_{ij}$ in the 
low-energy effective Lagrangian look different for different choices of 
basis, but this is simply due to field redefinitions---observables are 
independent the choice of basis.  Thus, when an ensemble of 
$\{K^a_{ij},\, \lambda^{ab}_{ij}\}$ is determined by considering vacua with 
various geometries and gauge-field configurations, the statistics of the 
flavor observables are basis independent.

\subsection{Gaussian Zero-mode Wavefunctions}
\label{ssec:Gaussianzeromodes}

If one considers generic internal manifolds $X$ and generic configurations 
of gauge-field backgrounds on the various $X$, then one expects generic 
Yukawa couplings to arise in the low-energy effective theory.  Given that 
we observe patterns in the flavor observables (such as those listed in the 
introduction), we expect our corner of the landscape to be described by 
some subset of the generic possibilities.  From a "bottom-up" 
perspective, one might first identify our corner of the landscape using 
phenomenological considerations, and only afterwards try to explain why 
this subset of vacua is preferred, whether it be by sheer statistics, 
dynamical effects, or anthropic selection.  While we do not attempt to 
identify all of the subsets of vacua in which our vacuum is typical, the 
present authors pointed out in Refs.~\cite{HSW1,HSW2} that the observed 
patterns of flavor are typical in at least one particular corner of the 
landscape.

If $X$ is a $T^m$-fibration on a $(D\!-\!m)$ dimensional base manifold $B$, 
and if the typical size of $T^m$ is smaller than that of $B$, then it is 
known that a Kaluza--Klein zero mode is effectively localized in a 
$(D\!-\!2m)$ dimensional subspace of $B$.  This localization can be 
understood intuitively as follows:  gauge fields tangential to the 
$T^m$-directions effectively become $m$ independent mass parameters for 
fields on $B$, and if the values of these ``mass parameters'' vary over $B$, 
then fermion zero modes are localized on $B$ where the mass parameters 
vanish---a mechanism known as a domain wall fermion.  Since there are $m$ 
independent mass parameters, the zero-mode fermions are localized in $m$ 
directions within the $(D\!-\!m)$ dimensional manifold $B$.  Wavefunction 
amplitudes decrease rapidly away from the locus of localization; the 
wavefunctions are approximately Gaussian in profile.\footnote{There is a 
phenomenological motivation to consider the localized gauge-field flux of 
$\U(1)_Y$~\cite{HallNomura} .  When such a $\U(1)_Y$ flux is localized in 
the internal space $X$, the zero-mode wavefunctions tend to have (often 
linear-exponential) hypercharge-dependent special behavior around the 
localized flux.  As a toy model, the Gaussian landscape below does not 
include this effect.  This is in part because the localized $\U(1)_Y$ flux 
is just one of many possible ways to break an $\SU(5)_{\rm GUT}$ symmetry.  
Although it may be possible to modify the Gaussian landscape to implement 
this mechanism, we do not do this here.}

As pointed out in Ref.~\cite{AS}, localized wavefunctions on extra dimensions 
easily generate hierarchically small Yukawa eigenvalues.  
Refs.~\cite{HSW1,HSW2} found this result to readily translate to the 
compactification scenario described above.  Furthermore, it was found (and 
will be elaborated upon below) that this picture explains the other major 
patterns of flavor, including the pairing structure and generation 
structure in the quark sector, and large lepton mixing.  
Thus, ``Gaussian landscapes'' are proposed as toy models of our corner 
of the string 
landscape.  Understanding Gaussian landscapes can shed light on the 
microscopic dynamics relevant to our corner of the landscape, and thereby
help address the question ``{\em why}'' of the introduction.

\subsection{Gaussian Landscapes}
\label{ssec:Gaussianlandscapes}

In practice, it is extremely technically involved to carry out the program
outlined in Section~\ref{ssec:KKreduction}.  One must first identify a 
stable gauge-field configuration on a compactified space $X$.  This 
involves solving a non-linear partial 
differential equation on a curved manifold.  One must then find the 
zero mode wavefunctions $\varphi^a_i(y)$; note that $\varphi^a(y)$ is a 
multi-component field, because it sits in a non-trivial representation of 
$\SO(D)$ as well as in a non-trivial gauge-field background.  Finally, an 
overlap integration must be performed over $X$ to obtain $K^a_{ij}$ and 
$\lambda^{ab}_{ij}$.  Note that one must calculate the metric on 
$X$ in order to perform this overlap integration, because the metric enters
Eqs.~(\ref{eq:gen-K}) and~(\ref{eq:gen-Yukawa}).  All this generates just 
one element of the landscape 
ensemble of $\{K^a_{ij},\,\lambda^{ab}_{ij}\}$; the process must be repeated 
many times to get a sense of the statistics of the landscape.  
Ref.~\cite{Douglas} takes a step toward performing such an analysis.

However, all of this may not be necessary to understand the observed
patterns of flavor.  If we may speak with the benefit of hindsight,  the 
distributions of flavor observables are quite broad across subsets of the 
landscape ensemble.  Our vacuum is selected randomly (modulo any anthropic 
or cosmological effects) from the landscape, so it is not important to 
understand observables with great precision.  Instead, we seek to 
understand the broad-brush patterns of flavor, assuming that the precise 
values of flavor observables are essentially accidents.

For this purpose, we introduce a number of simplifying assumptions, hoping 
to capture the essence of flavor in a certain corner of the string 
landscape, while setting 
aside the complicating details.  First, we use only the base manifold $B$ 
when calculating overlap integrals to construct $K^a_{ij}$ and 
$\lambda^{ab}_{ij}$.  Second, although the zero-mode wavefunctions 
$\varphi^a$ are multi-component fields, we treat them as if they were single 
component fields.  Since we consider the approximately Gaussian profiles of 
zero-mode wavefunctions to be the origin of hierarchical Yukawa couplings, 
we expect single component wavefunctions to carry the most important 
information.  For concreteness, we use wavefunctions with strictly Gaussian 
form (made periodic on $B$); e.g. for the $i$th zero mode of the 
representation $a$, 
\bea
\varphi^a_i(y) \sim \exp\left[-\frac{1+ir_a}{2d_a^2}
\left(y-y^{a_i}\right)^2\right] \,.
\label{wvfcn}
\eea  
The peak position $y^{a_i}$ is a point in $B$, the width $d_a$ determines 
the degree of localization of the Gaussian profile, and $r_a$ is a possible 
complex phase that we include for later reference.\footnote{The inclusion
of the complex phase $r_a$ is not without motivation; it is an approximate 
form for zero-mode wavefunctions on a torus with generic complex structure
(see Section 7 of Ref.~\cite{HSW2} for more details).  However, it is 
unclear how complex phases enter the zero-mode wavefunctions when $B$ is 
not a torus.  We introduce complex phases into zero-mode wavefunctions in
this way merely to explore a possibility.}  
The normalization of these wavefunctions is arbitrary (it does not affect 
observables).

To establish a more convenient notation for later, we write the relevant 
terms of the low-energy Lagrangian,
\bea
\mL &\supset&  K^a_{ij}\,\bar{a}_ii\gamma^\mu\! D_\mu\, a_j 
+ \lambda^u_{ij}\, \bar{u}_i\, q_j \, h 
+ \lambda^d_{ij} \, \bar{d}_i\, q_j \, h^\dagger \nn\\
&& +\,\, \lambda^e_{ij} \, \bar{e}_i\, l_j \, h^\dagger 
+\lambda^n_{ij} \, \bar{n}_i\, l_j \, h +
\lambda^M_{ij} \, \bar{n}_i\, \bar{n}_j \, \phi \,,
\label{L}
\eea
where $n_i$ denotes a (heavy) right-handed neutrino.  The indices $i,j$ 
label generation, but note that we do not assume three ``generations'' of 
right-handed neutrinos.  The kinetic matrices $K^a$ are given by overlap 
integration of zero-modes, c.f. Eq.~(\ref{eq:gen-K}), here written 
\bea
K^a_{ij} = M_*^D\!\int_B d^Dy\sqrt{g}\;
{\varphi^a_i}^*(y) \, \varphi^a_j(y) \,. 
\label{KI}
\eea 
The up-type Yukawa matrix, c.f. Eq.~(\ref{eq:gen-Yukawa}), is given by
\bea
\lambda^u_{ij} = g_* M_*^D\!\int_B d^Dy\sqrt{g}\; p(y)\;
\varphi^{{\bar u}}_i(y) \, \varphi^q_j(y) \,
\varphi^{h}(y) \,,
\label{overlap}
\eea
where for later reference we have introduced a phase function $p(y)$ that 
might arise due to the ``$\Gamma$-structure'' in $\Gamma^m\varphi^h_m$ of 
Eq.~(\ref{eq:gen-Yukawa}).\footnote{To be more precise, we can
write $\Gamma^m=\Gamma^p e^m_p$, where the $\Gamma^p$ are fixed-value 
matrices, like the Pauli matrices, that satisfy the algebra of the internal
$\SO(D)$.  Just like with the Pauli matrices, the $\Gamma^p$ may have 
mutually different complex phases.  $e_p^m$ is a vierbein on the 
internal manifold, and thus depends on $y$.  Therefore, we expect a 
$y$-dependent complex phase in $\Gamma^m$.  Our introduction of a phase
function $p(y)$ is an attempt at modeling the possible effects of this.}   
The Yukawa matrices $\lambda^d_{ij}$, $\lambda^e_{ij}$, and $\lambda^n_{ij}$
are all defined analogously.  Although the wavefunction of the Higgs boson 
will in general have complicated structure, in the Gaussian landscape we 
treat it as a single-component wavefunction with Gaussian 
profile.\footnote{\label{fn4}Note, however, that there may be important 
consequences of the wavefunction $\varphi^h_m(y)$ (and $\varphi^h_m(y)^*$)
having multiple components.  Suppose that $G \supset G' \times G_{SM}$, and 
that the $G'$ symmetry is broken by the gauge-field background.  Then the 
$\varphi^a_i(y)$ are all zero modes under the gauge-field background in 
$G'$, each in a certain representation $\rho_a$.  Suppose also that 
$\varphi^h_m(y)$ is in a representation $\rho_h$, and $\varphi^h_m(y)^*$ 
in $\rho_h^\times$.  Then the up-type Yukawa couplings are given by overlap 
integration of trivial components in the irreducible decomposition of 
$\rho_{a=\bar{u}} \otimes\rho_{a=q} \otimes \rho_h$, while the down-type 
Yukawa couplings by that of trivial components of 
$\rho_{\bar{d}} \otimes \rho_q \otimes \rho_h^\times$.  Thus, a single 
component wavefunction $\varphi^h(y)$ for $\lambda^u$ and $\lambda^\nu$ and 
precisely its complex conjugate wavefunction $\varphi^h(y)^*$ for 
$\lambda^d$ and $\lambda^e$ may introduce too restrictive a relation between 
the Yukawa couplings in the different sectors.  To account for this 
issue, one may extend the Gaussian landscape by introducing additional 
parameters and/or by scanning the ``zero-mode wavefunction" of the Higgs 
boson.}
This is reasonable, since linear changes in the gauge-field 
background result in quadratic changes in the mass-square parameter in the 
quadratic mode equation of the bosonic zero modes;  note that 
the ground-state wavefunction of a harmonic oscillator is Gaussian.

We assume low-energy left-handed neutrino masses arise from the seesaw 
mechanism.  Hence these masses are determined by integrating out 
the right-handed neutrinos, giving the effective interactions
\bea
\left(C^\nu_{ij}/\vev{\phi}\right)\, \ell_i\,\ell_j\,h\,h\,, 
\quad {\rm where} \quad
C^\nu_{ij} = 
\left( \lambda_n^T \, \lambda_M^{-1} \, \lambda_n \right)_{ij} \,.\,\,\,\,
\label{seesaw}
\eea  
The field $\phi$ is a scalar that couples to the $n_i$ in the Majorana 
mass term because the Standard Model gauge singlets $n_i$ are not 
necessarily gauge singlets in the higher-energy gauge group.  In general 
there may be more than one scalar $\phi$, and we take $\varphi^\phi$ to 
represent the effective wavefunction of the set of scalars that generate a 
mass term for the $n_i$, after acquiring an effective vacuum expectation 
value $\vev{\phi}$.

We assume the Majorana mass matrix $\lambda^M$ of the right-handed 
neutrinos $\bar{n}_i$ is generated randomly, via overlap integration in 
analogy to $\lambda^u$ and the other Yukawa matrices.  There is more 
than one possible origin of the right-handed neutrino mass interactions, 
and the overlap integration for $\lambda^M$ may not have the same phase 
functions as $\lambda^{u,d,e,n}$ coming from the $\Gamma^m$.  Thus, we 
drop $p(y)$ for $\lambda^M$ in the numerical simulations of 
Section~\ref{sec:numerical}.  In fact, it is possible that neutrino 
masses are determined quite differently than this, however our assumptions 
capture the spirit of the Gaussian landscape, which replaces flavor 
symmetry charges with overlap integration of localized wavefunctions.

In order to generate an ensemble of vacua and statistical distributions 
of the flavor observables, we scan each of the $y^a_i$ randomly and 
independently over $B$.  This ansatz borrows a hint from the fact that 
instanton center coordinates can be chosen freely in multi-instanton 
configurations, and fermion zero modes are localized around isolated 
instanton centers.  It should be emphasized, however, that this intuitive 
ansatz is far from justified from the top-down perspective outlined 
above---further understanding is necessary to develop a rigorous 
connection between microscopic theoretical formulations and this ansatz
for generating statistics of flavor observables.  Nevertheless, we 
consider it worthwhile to study whether such an ansatz generates 
phenomenologically acceptable distributions of flavor observables or 
not, before expending too much effort on rigorous calculations.  This 
simplified toy model of the string landscape is what we refer to as a 
Gaussian landscape.

The Gaussian landscape above differs from that of Refs.~\cite{HSW1,HSW2} 
in the inclusion of the kinetic overlap integrals, Eq.~(\ref{KI}).  These
were ignored in the toy model of Refs.~\cite{HSW1,HSW2} because it was 
guessed that they would not significantly affect the essence of the results.  
We here find, however, that they have important implications for obtaining 
large lepton mixing in the Gaussian landscape.


\section{Numerical Simulation of the Gaussian Landscape}
\label{sec:numerical}

An analytic understanding of the Gaussian landscape is made difficult 
in part by the complicated set of operations necessary to convert the 
theoretical predictions for the $K^a_{ij}$ and $\lambda^a_{ij}$ into 
predictions for the observable masses, mixing angles, and CP-violating
phases.  On the other hand, it is straightforward to perform such 
computations numerically.  Thus, we study the Gaussian landscape using
a numerical simulation, generating a large ensemble of sets of flavor 
parameters in order to represent their landscape distributions.\footnote{We 
are grateful to Nathan Moore for helping to improve the efficiency of our 
numerical algorithm.}  
As described in Section~\ref{sec:background}, we populate the landscape by
taking the various peak positions $y^{a_i}$ to scan randomly and 
independently over the internal manifold.  Meanwhile, we take the 
other parameters---the microscopic gauge coupling $g_*$, cutoff scale 
$(M_*L)^D$, widths $d_a/L$, phases $r_a$ and phase function $p(y)$, the 
number of right-handed neutrinos $N_n$, and even the internal manifold 
$B$---as given.      

Some of the parameters that we take as given may in fact be uniquely 
determined by the microscopic theory, while others will scan in the full 
string landscape.  Our strategy can be seen as focusing attention on a 
certain slice of the landscape.  This creates a handle for understanding 
the results, and meanwhile avoids the problem of guessing how these 
parameters are distributed among vacua in the landscape.  Although the 
choice of values for these parameters is arbitrary, this will not be a 
problem if the values we choose are not atypical in the full string 
landscape.  For example, we choose $d_u/L$ so as to match the observed 
level of hierarchy between up and top quark masses.  It could be the 
microscopic theory uniquely determines this width, in which case our 
choice can ultimately be checked against this prediction, or that 
anthropic selection is important in determining the hierarchy, in which 
case our choice automatically includes this effect, or that $d_u/L$ scans 
with a broad distribution that includes the value 
we choose, in which case the observed level of hierarchy is simply an 
accident.  As is indicated below, for many changes in the values of these 
fixed parameters, the patterns of flavor remain intact.

Let us now describe our choices for these parameters.  First, 
we choose $D=2$ extra dimensions, taking their geometry to be that of the 
square torus $T^2$, each dimension having length $L$.  As is shown
in some detail in Ref.~\cite{HSW2}, the distributions of flavor observables
in Gaussian landscapes are broadly independent of the geometry of the 
manifold $B$, while the number of extra dimensions $D$ primarily affects 
a single detail of these distributions, involving the distribution of the 
most massive particle in each generation.  In the latter case, $D=1$ stands
out from the others, so we expect our choice of geometry to closely describe 
the results of any choice $D\geq 2$ on any smooth geometry.  Because it is 
unclear how CP violation should enter the Gaussian landscape---for example 
via the phase function $p(y)$ in Eq.~(\ref{overlap}), via wavefunction phases 
$r_a$ in Eq.~(\ref{wvfcn}), or by some other mechanism---we set the issue 
aside for the moment and focus on the CP-conserving flavor observables, 
choosing $p(y)=1$ and $r_a=0$ for all $a$.  We set 
$g_{\rm eff}\equiv g_*/M_*L=0.5$ to visually match the observed $t$, $b$, and 
$\tau$ Yukawa couplings; increasing or decreasing $g_{\rm eff}$ merely shifts 
the set of fermion mass distributions to larger or smaller values (note that
$g_{\rm eff}$ may determine the gauge couplings of the Standard Model, and 
therefore anthropic effects may be important in selecting its value).  

We choose $d_a/L\equiv d/L=0.1$ for all fields except the 
down-type quark singlets and the lepton doublets, for which 
$d_d/L=d_\ell/L\equiv d_{\bf \bar{5}}/L=0.3$.  The choice $d/L=0.1$ is 
important for setting the observed level of mass hierarchy between the 
heaviest and lightest generations.  The logarithmic spacing between typical 
masses in these generations is proportional to $(L/d)^2$~\cite{HSW2}; the 
level of hierarchy that we observe can be seen as an accident or as 
anthropically selected to generate appropriate atomic structure (see for
example Refs.~\cite{anth}).  Meanwhile, 
in order to obtain large lepton mixing, the lepton doublet width $d_\ell$ 
must be larger.  Hence we set $d_\ell/L=0.3$.  We choose $d_{d}/L=0.3$ simply 
because we consider it plausible that these fields should have the same 
width, as they sit in the same $\SU(5)$ representation, 
${\bf \bar{5}}=\{\bar{d},\ell\}$.  As we shall explain, it would not hurt 
the fit to observation to give down-type quarks the same width as all of 
the other fields.      

Finally, we assume there are $N_n=20$ heavy right-handed neutrinos.  This 
number is chosen mainly for concreteness.  As we shall describe, $N_n=3$
does not provide a terrible fit to existing observations for some choices 
of model parameters, including those described above.  However, the fit to 
observation improves, and become more robust to changing the values of 
other model parameters, when $N_n$ is increased.  

\begin{figure*}[t!]
\begin{center}
\begin{tabular}{c c c c} 
\includegraphics[width=0.21\linewidth]{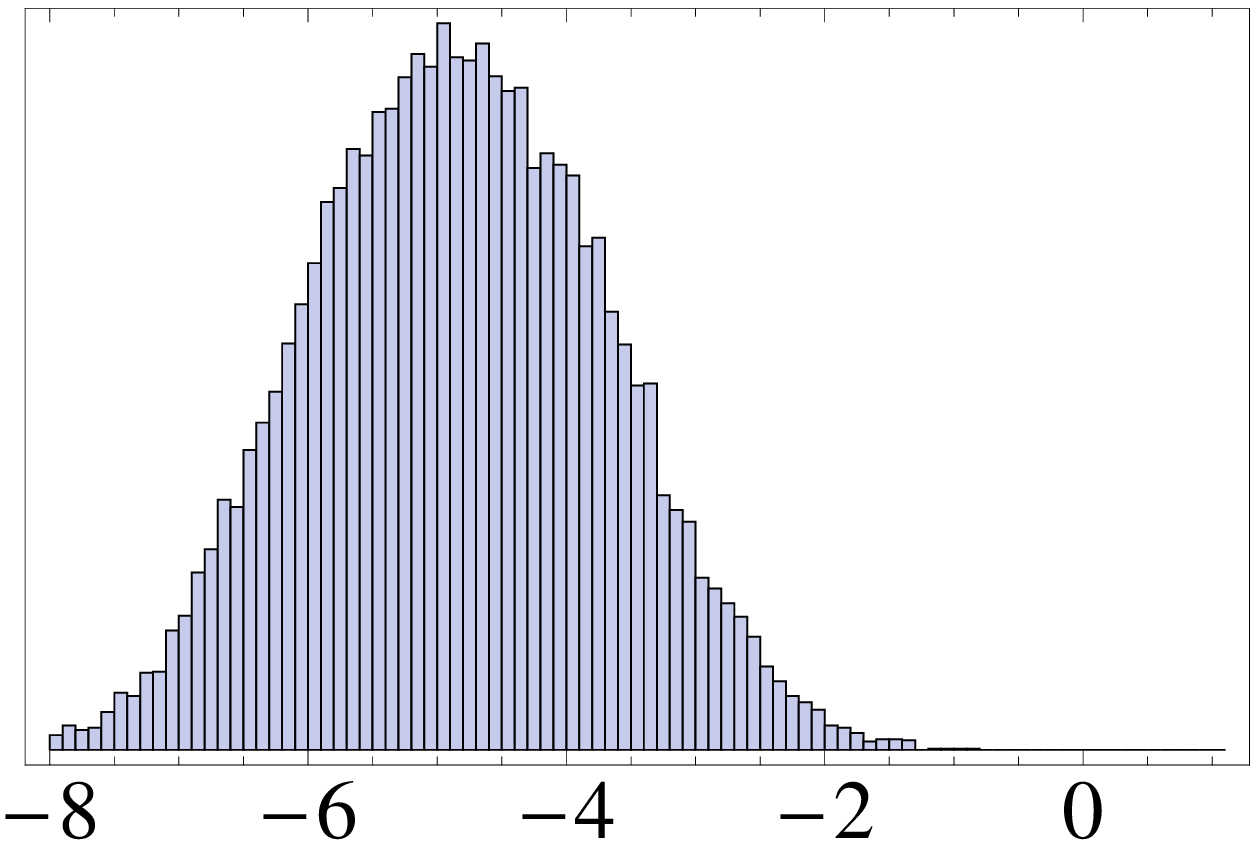} &
\includegraphics[width=0.21\linewidth]{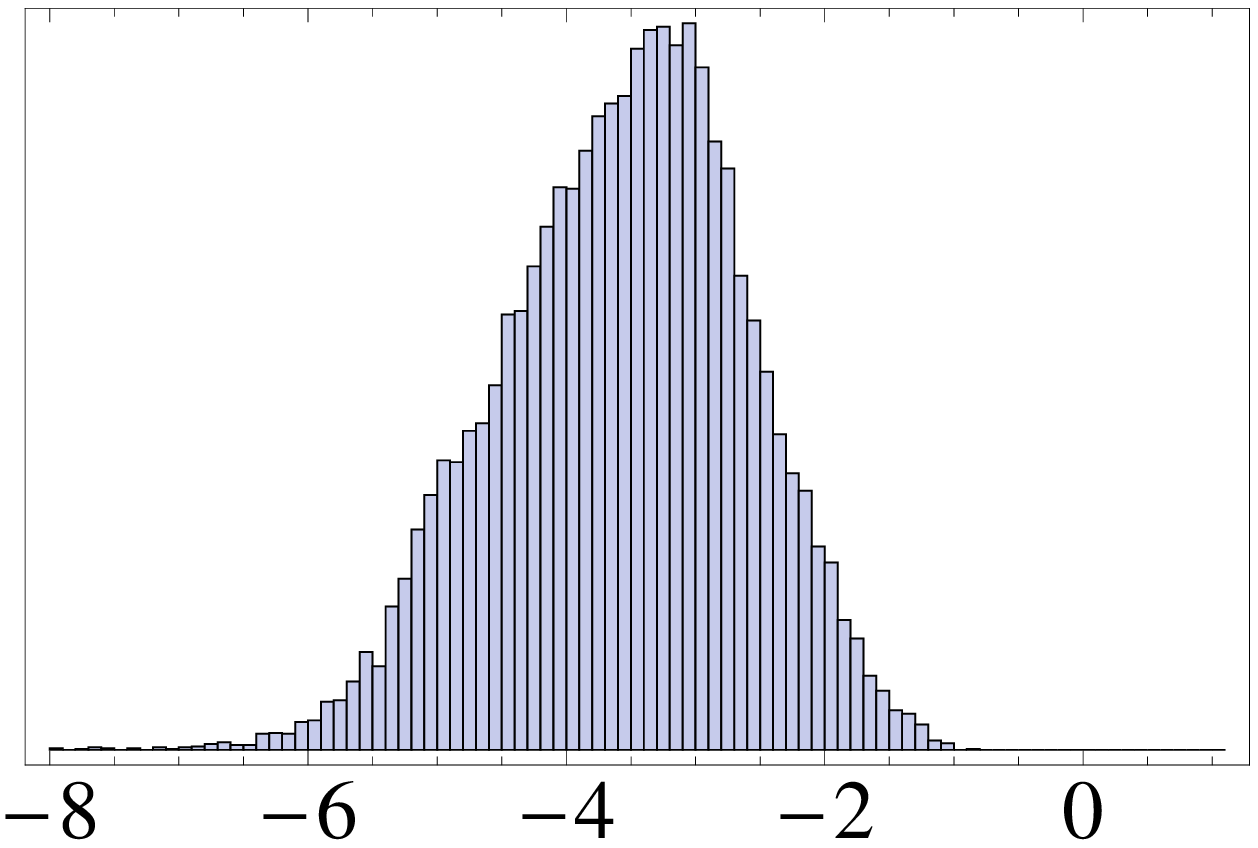} &
\includegraphics[width=0.21\linewidth]{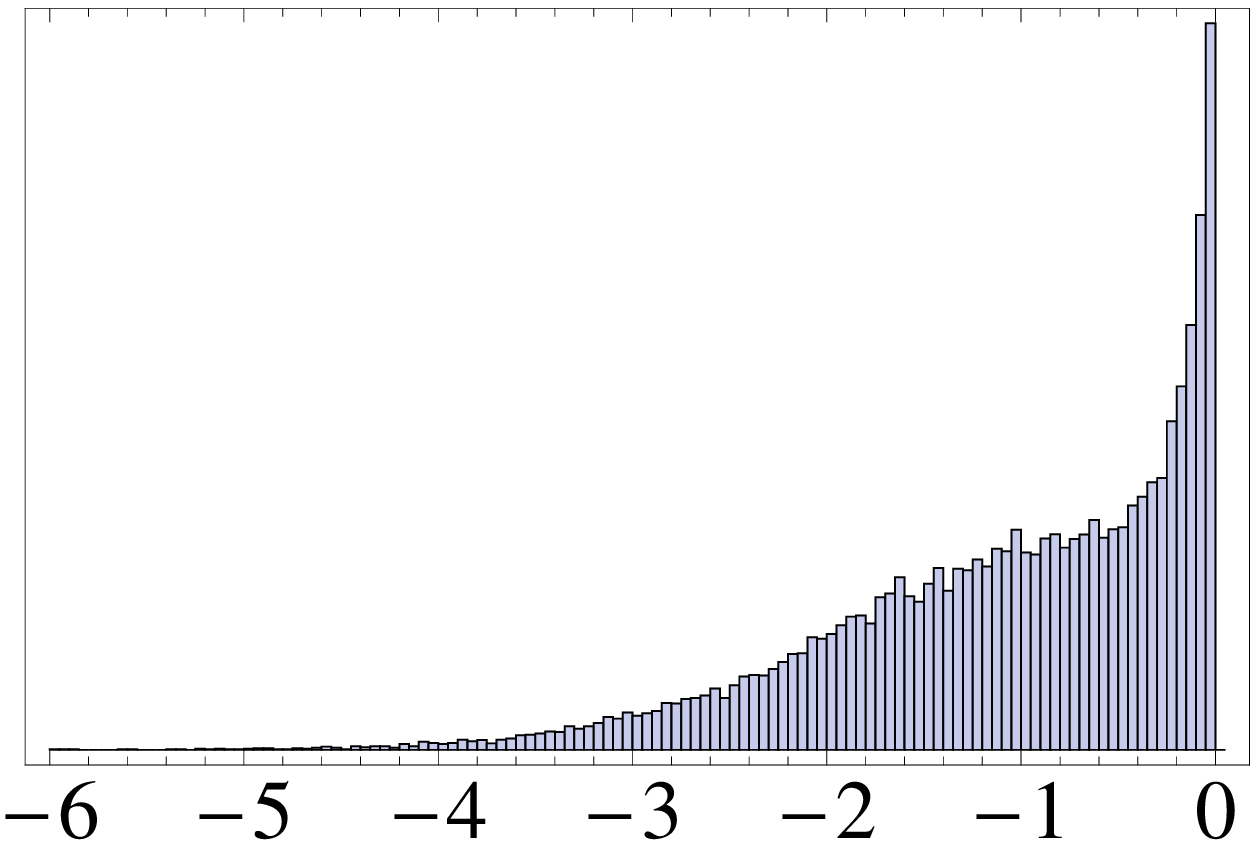} &
\includegraphics[width=0.21\linewidth]{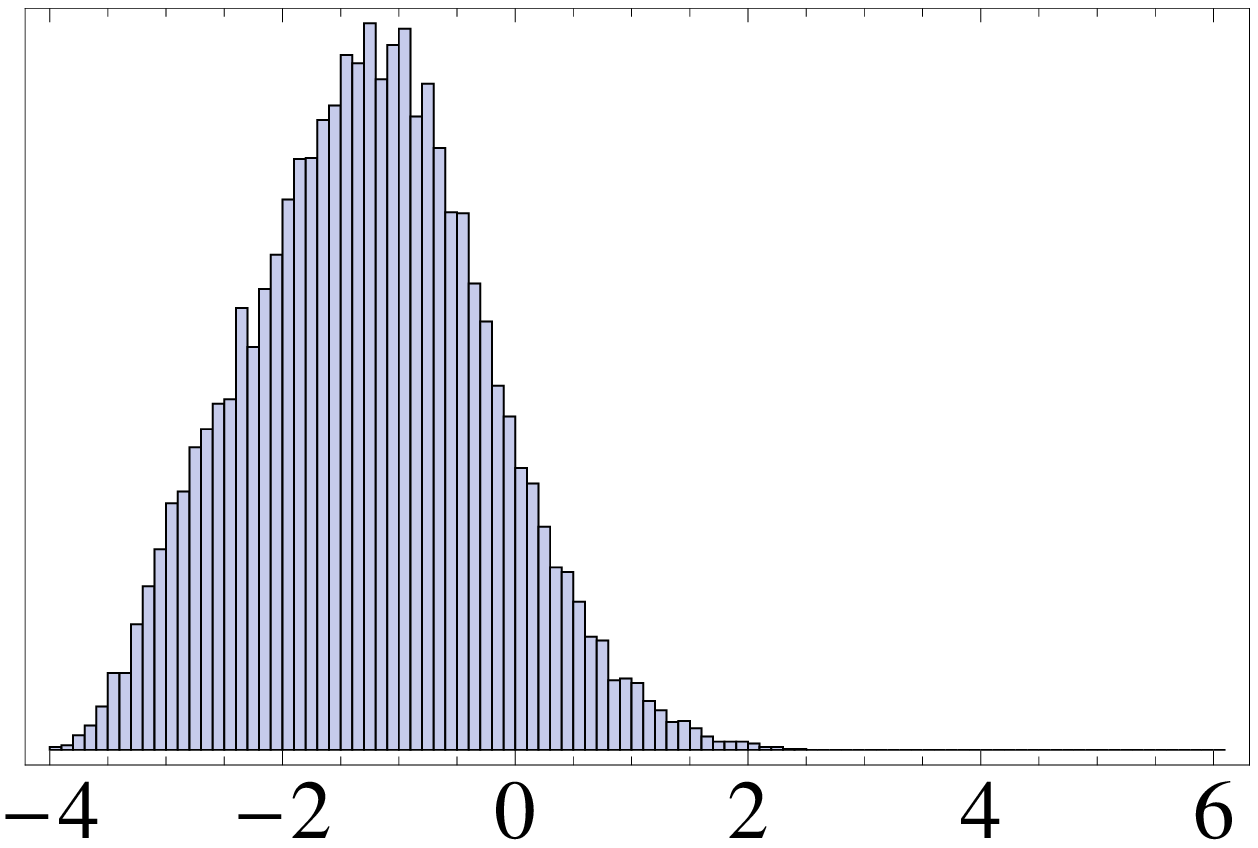} \\
$\log\lambda_{u}$ [$-5.5$] & 
$\log\lambda_{d,e}$  [$-5.2,-5.5$] &
$\log(2\theta^{\rm CKM}_{12}/\pi)$ [$-0.8$] &
$\log C^\nu_1$ \\
\includegraphics[width=0.21\linewidth]{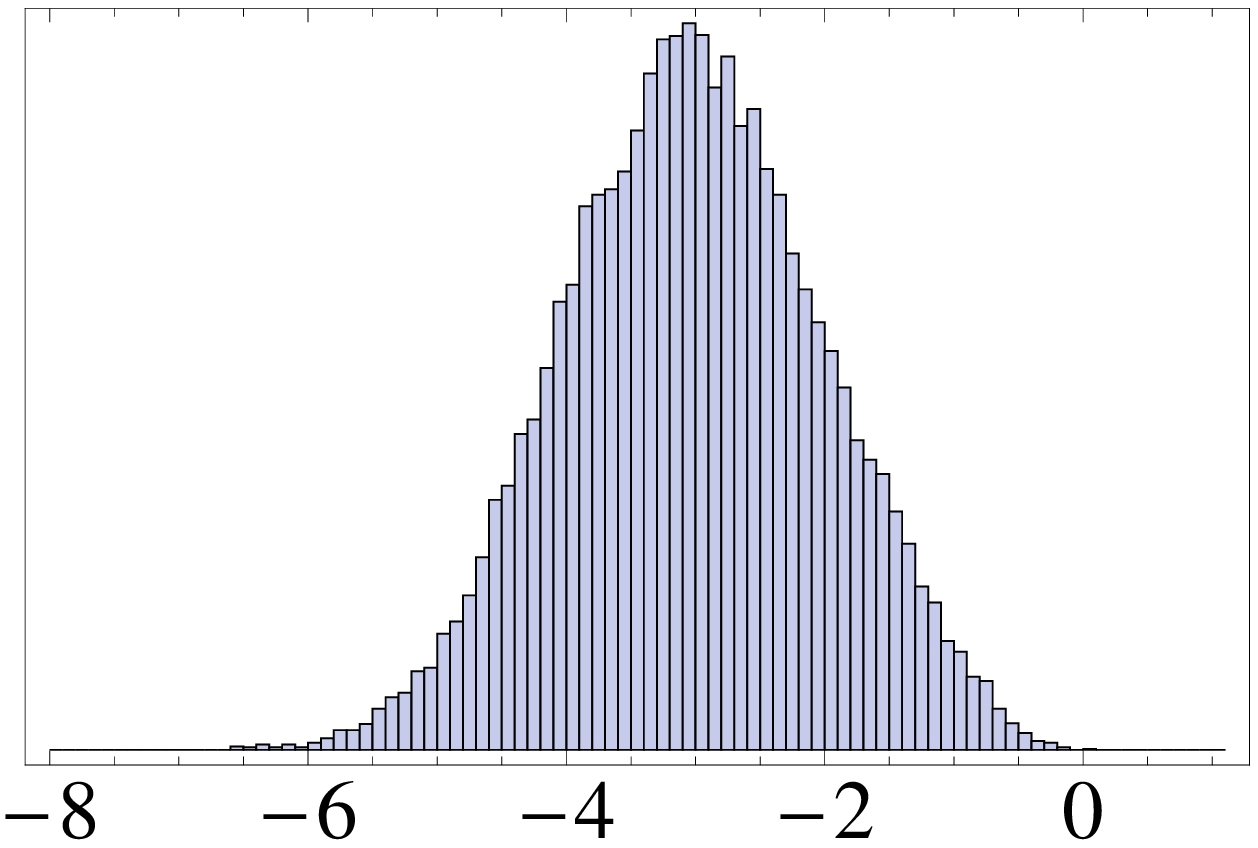} &
\includegraphics[width=0.21\linewidth]{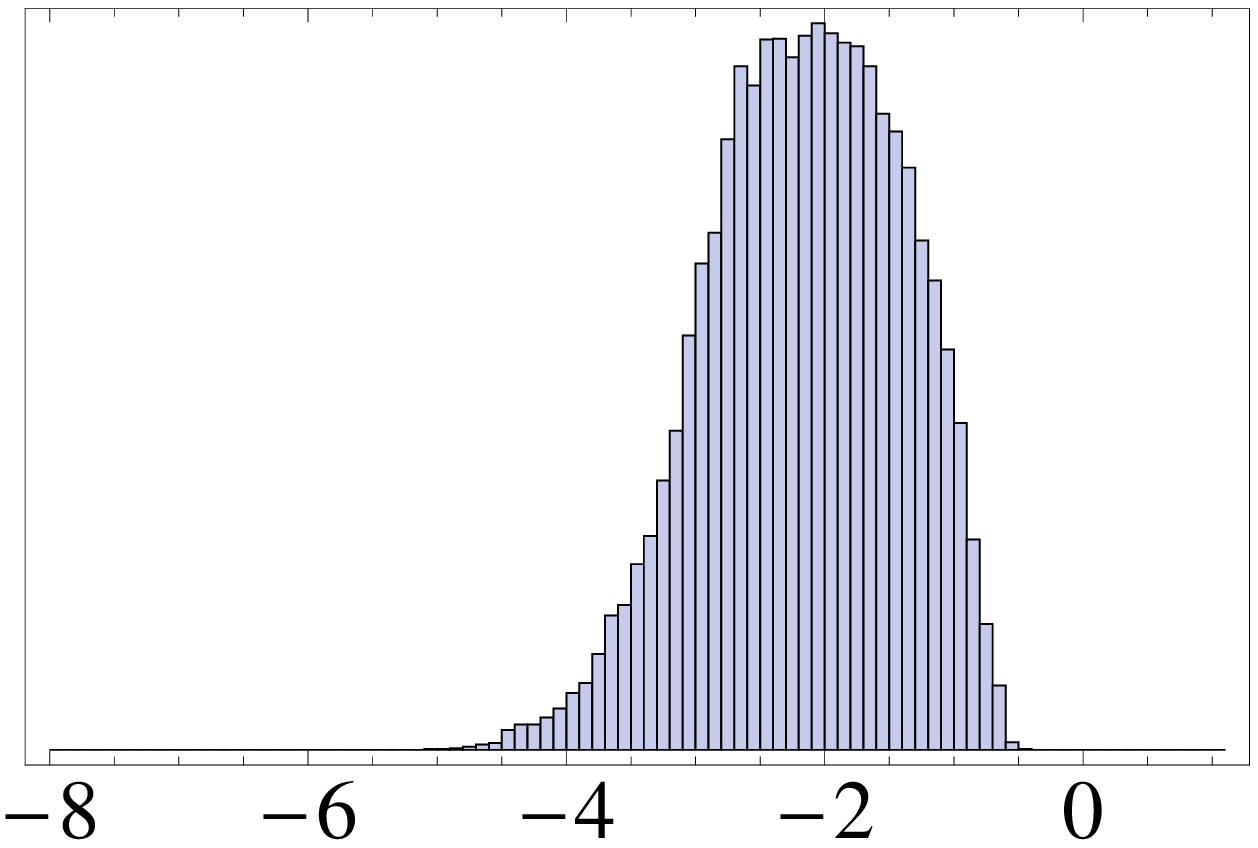} &
\includegraphics[width=0.21\linewidth]{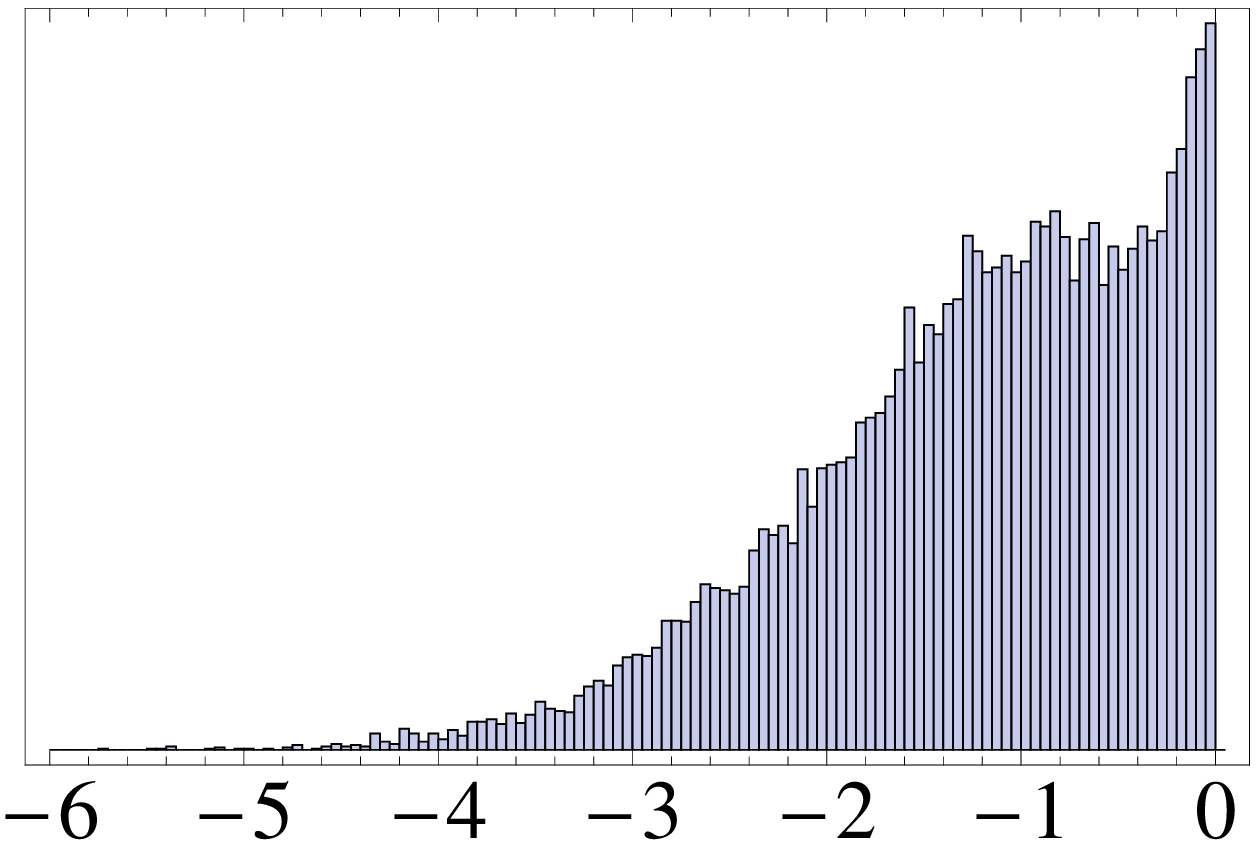} &
\includegraphics[width=0.21\linewidth]{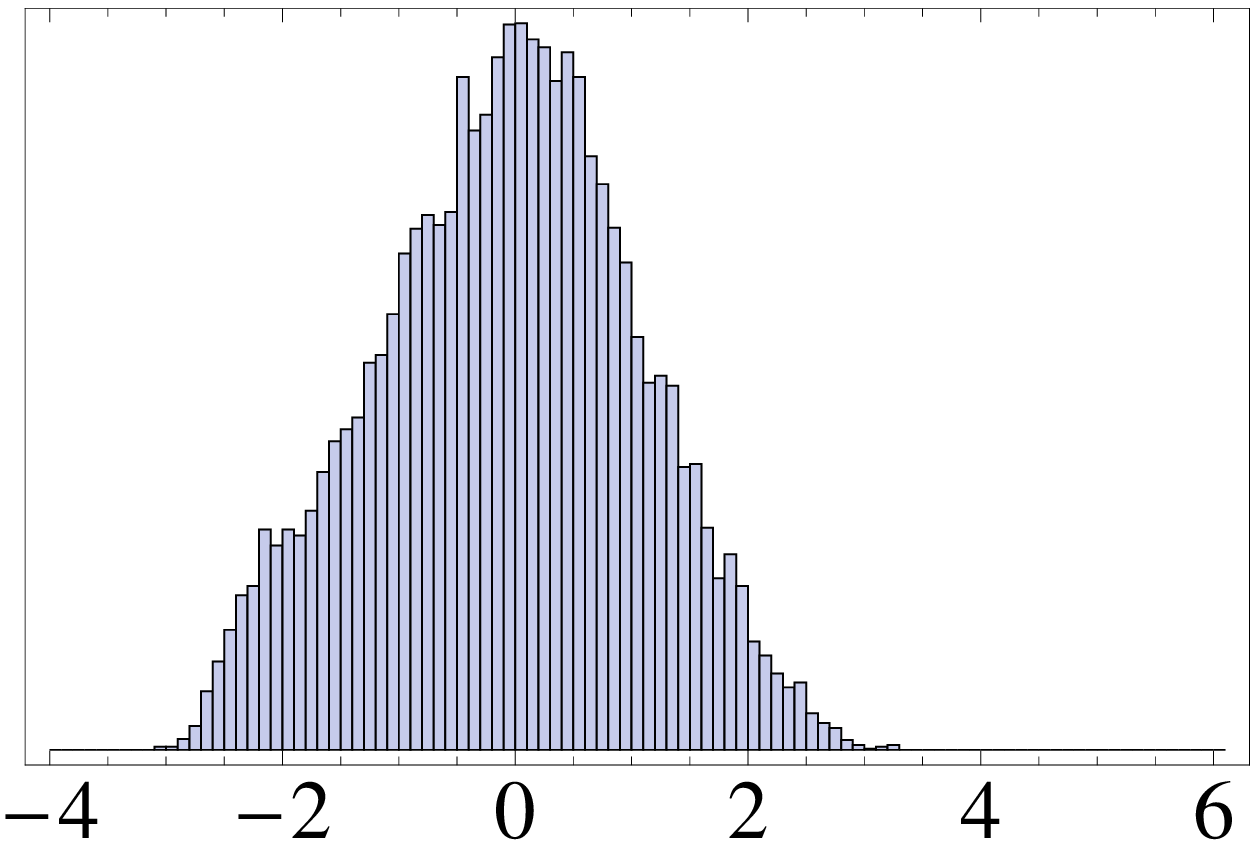} \\
$\log\lambda_{c}$ [$-2.9$] & 
$\log\lambda_{s,\mu}$ [$-3.9,-3.2$] & 
$\log(2\theta^{\rm CKM}_{23}/\pi)$ [$-1.6$] &
$\log C^\nu_2$ \\
\includegraphics[width=0.21\linewidth]{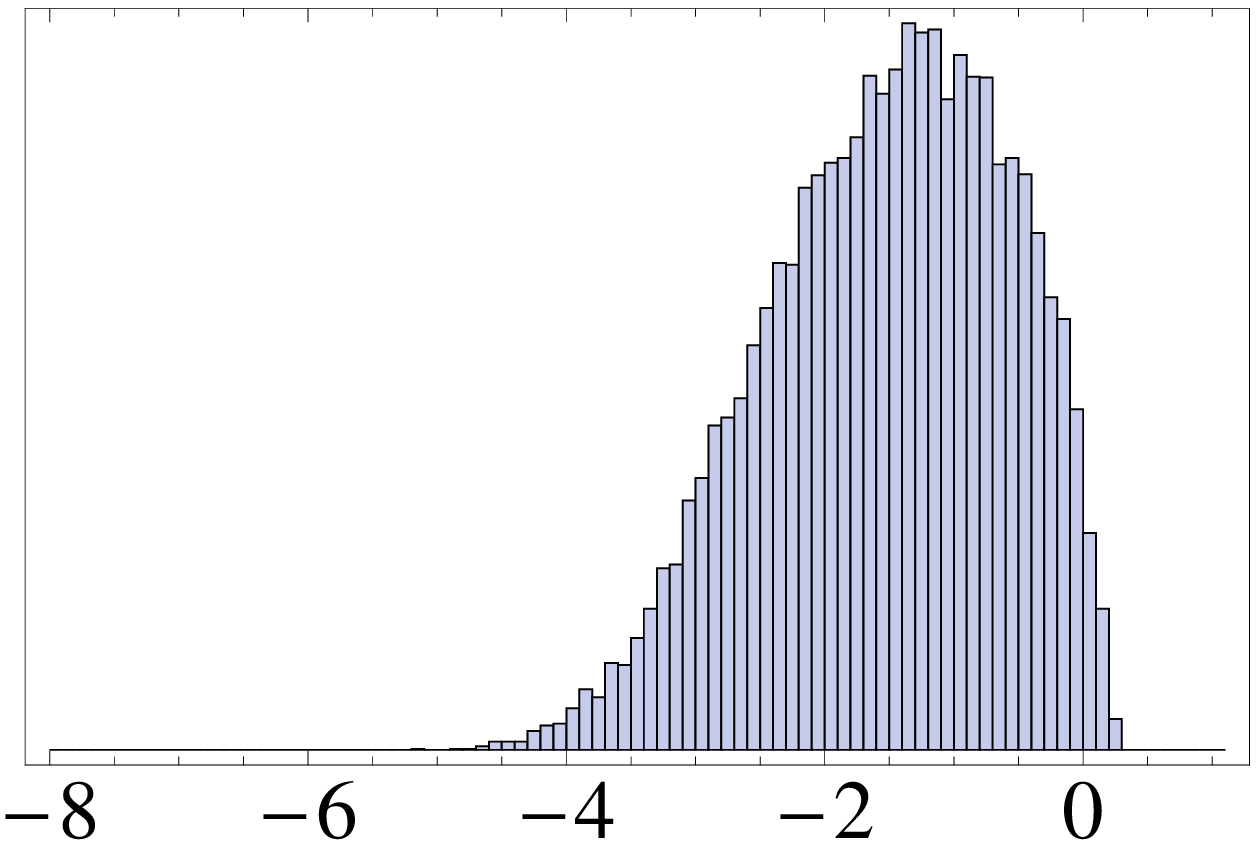} & 
\includegraphics[width=0.21\linewidth]{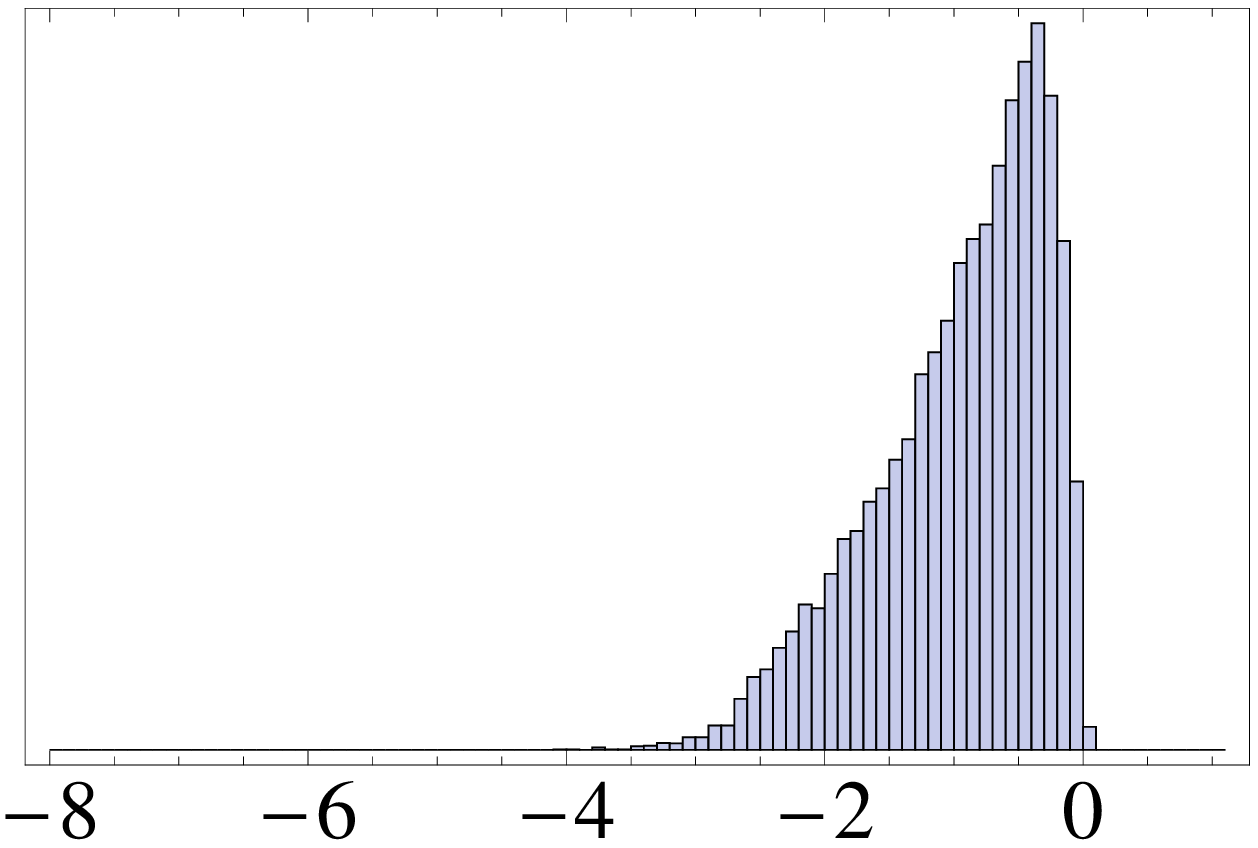} & 
\includegraphics[width=0.21\linewidth]{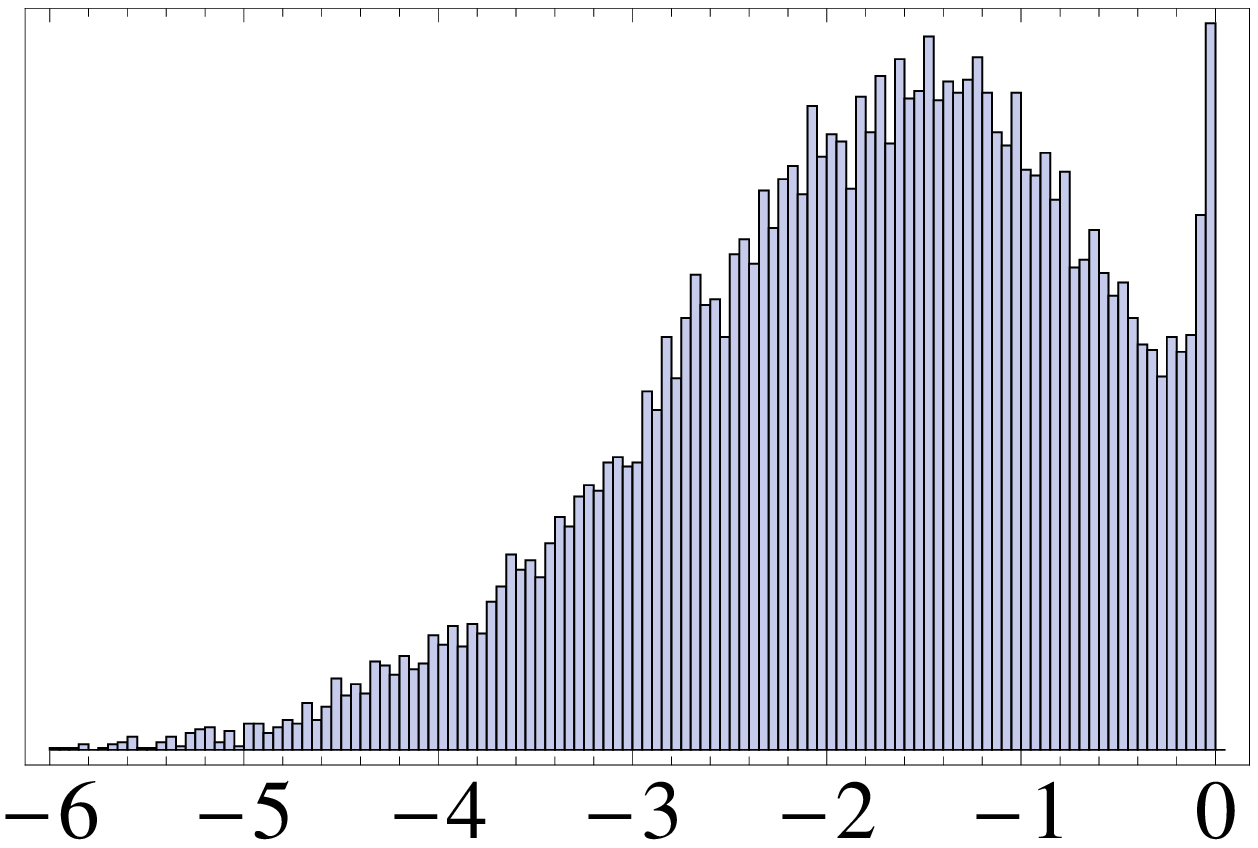} &
\includegraphics[width=0.21\linewidth]{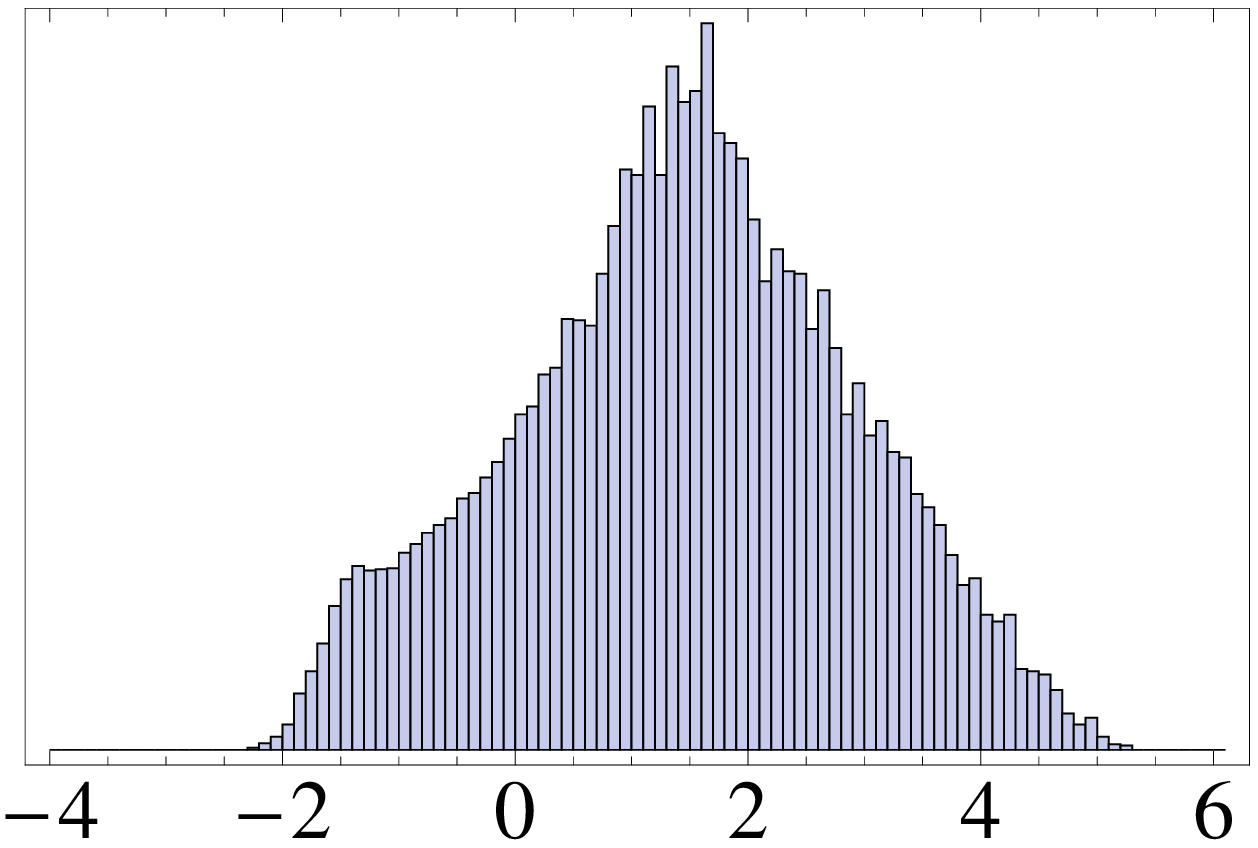} \\
$\log\lambda_{t}$  [$-0.3$] & 
$\log\lambda_{b,\tau}$  [$-2.2,-2.0$] & 
$\log\sin(\theta^{\rm CKM}_{13})$ [$-2.4$] &
$\log C^\nu_3$ \\
\includegraphics[width=0.21\linewidth]{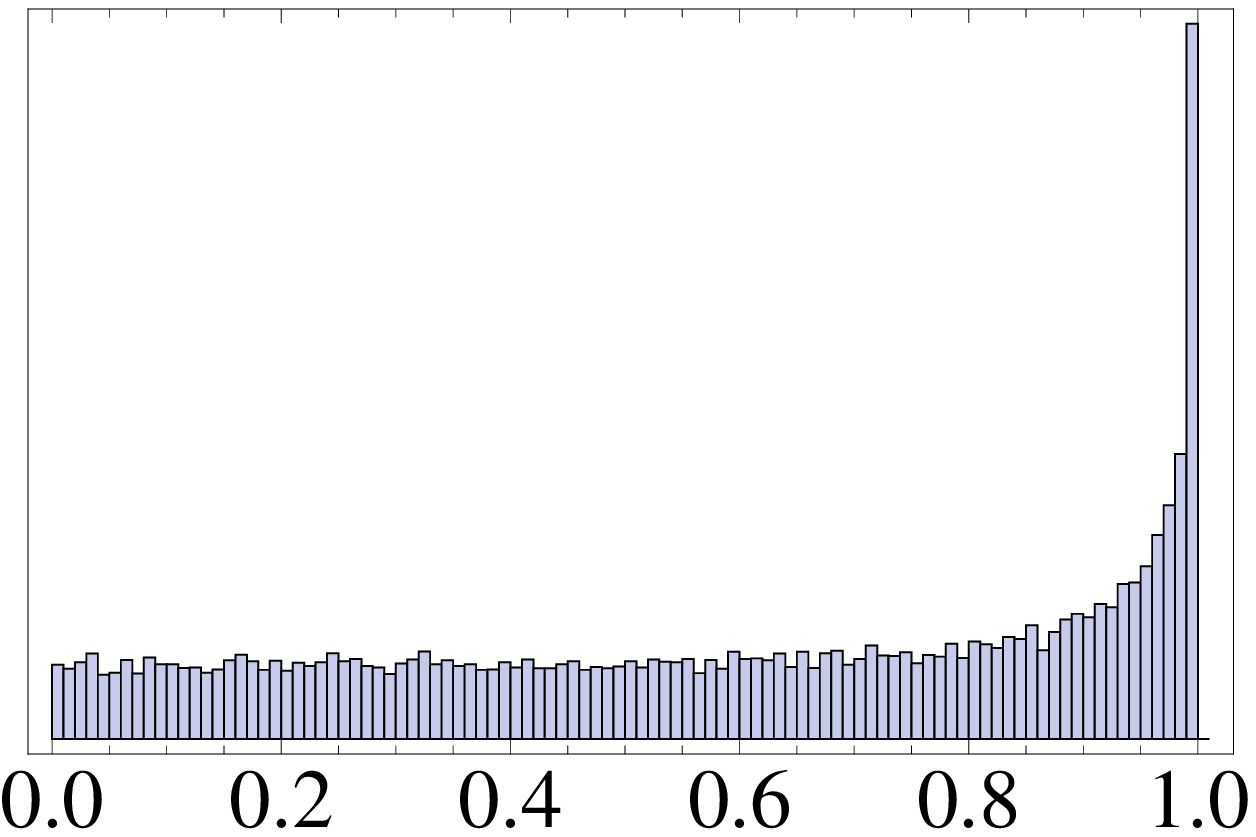} &
\includegraphics[width=0.21\linewidth]{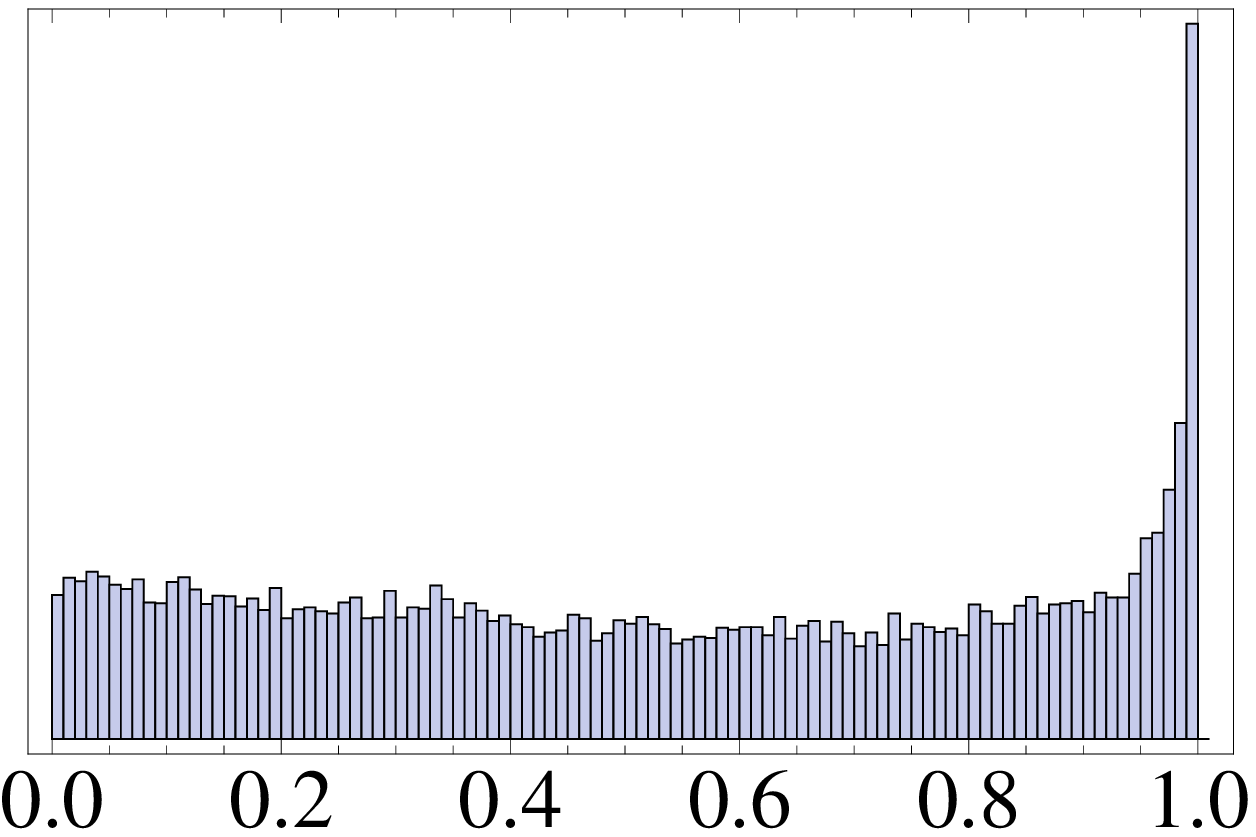} &
\includegraphics[width=0.21\linewidth]{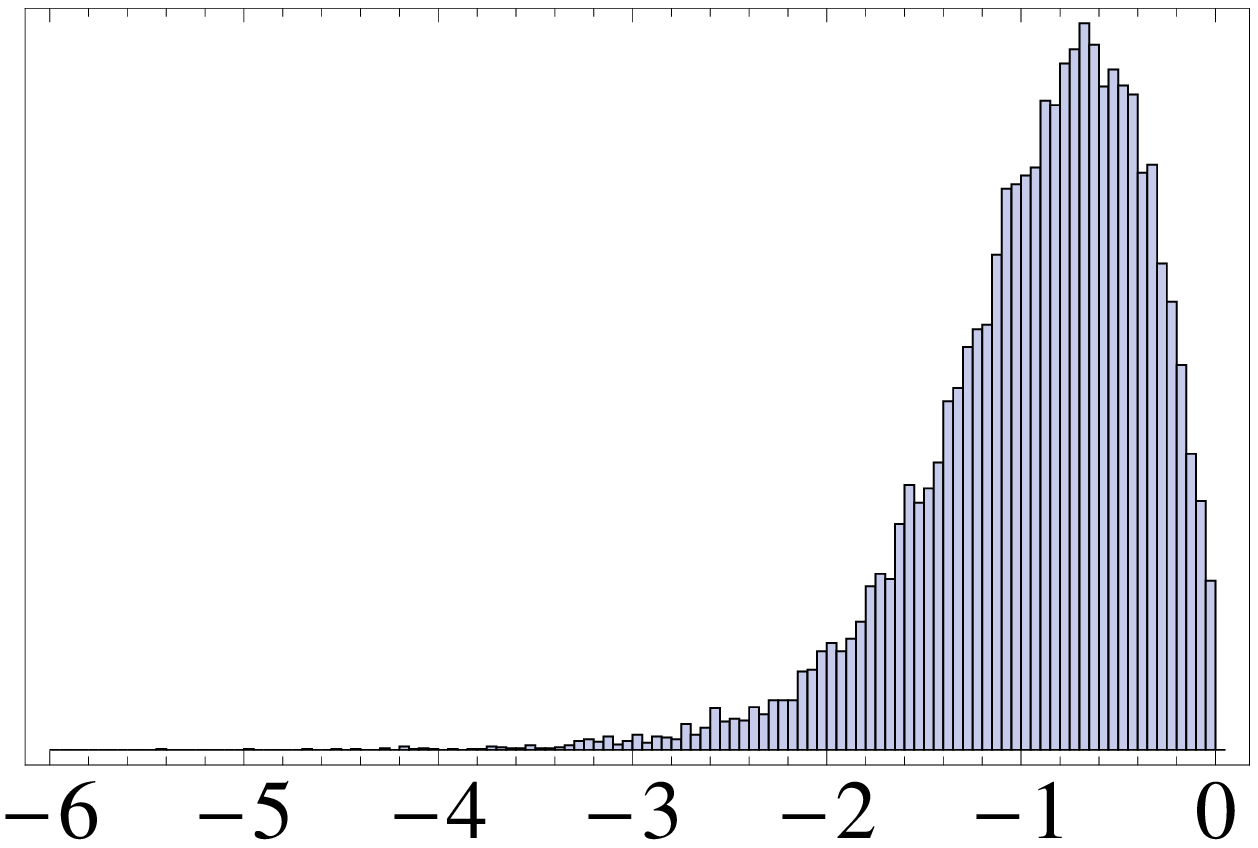} &
\includegraphics[width=0.21\linewidth]{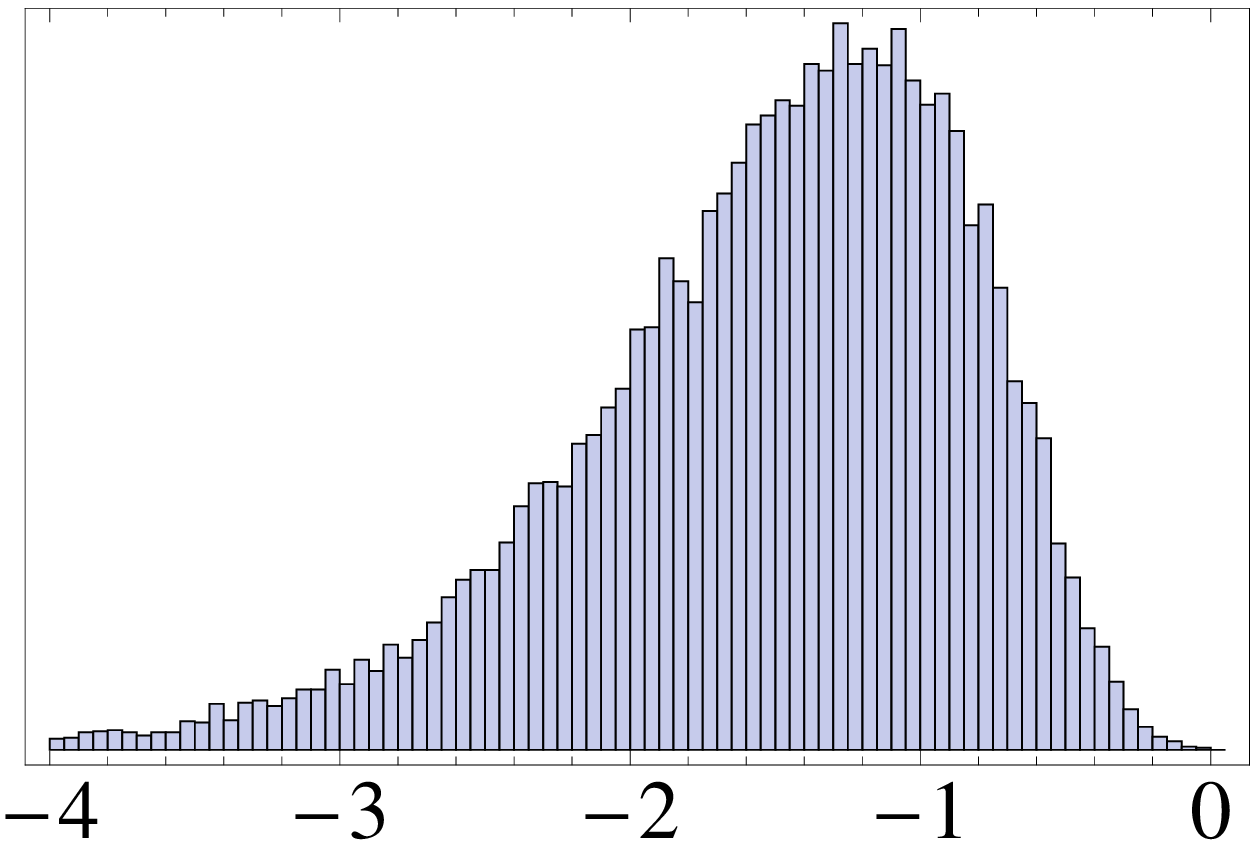} \\
$\sin(2\theta^{\rm PMNS}_{12})$ [$0.92\pm0.01$] &
$\sin(2\theta^{\rm PMNS}_{23})$ [$>0.96$] &
$\log\sin(\theta^{\rm PMNS}_{13})$ [$<-0.74$] &
$\log(m^\nu_2/m^\nu_3)$ [$-0.75\pm0.05$] \\
\end{tabular}
\caption{\label{fig:1}  Distributions of flavor observables for the 
Gaussian landscape on $T^2$.  We set $p(y)=1$, $r_a=0$ for all $a$, 
$g_{\rm eff}=0.5$, $d/L=0.1$, $d_{\rm \bar{5}}/L=0.3$, and assume $N_n = 20$ 
right-handed neutrinos.  All logarithms are base ten, $\lambda_a$ denotes 
the Yukawa coupling determining the mass of particle $a$, and $C^\nu_i$ is 
the $i$th eigenvalue of the low-energy neutrino mass matrix $C^\nu$; other 
notations conform to those of the Particle Data Group.  The scale 
$\vev{\phi}$ has been set so that the median of $C^\nu_2$ is one.  The 
numbers in brackets represent experimental values~\cite{PDG}, run up to the 
Planck scale.  See the main text for details.}
\end{center}
\end{figure*}

As we have already described, the peak positions $y^{a_i}$ are taken to
scan randomly and independently over the interval $[0,L]$.  Although 
there are 22 flavor observables, for $D\geq 2$ the number of scanning 
parameters exceeds this, and the Gaussian landscape makes no ``hard'' 
predictions.  Instead, one obtains (correlated) probability distributions
for the flavor observables.  These are reflected in the binned results of
a numerical simulation, displayed in Fig.~\ref{fig:1}.  For reference, 
Fig.~\ref{fig:1} also gives experimental values in brackets.  These have 
been obtained by using renormalization group flow to run Particle Data 
Group~\cite{PDG} mean values or constraints up to the Planck scale, 
where, for concreteness, we assume no particle content beyond the Standard 
Model and ignore the effect of the right-handed neutrinos (alternatively, 
one could extend the Gaussian landscape to cover low-energy supersymmetry).  
We here emphasize that Fig.~\ref{fig:1} displays the {\em statistical} 
distributions of flavor observables among vacua in the Gaussian landscape, 
and does not include cosmological or anthropic selection effects, the 
latter of which we anticipate would modulate at least the distributions 
of $\lambda_{u,d,e}$.    

The distributions in Fig.~\ref{fig:1} are all quite broad.  This is 
undesirable, from the point of view of experimental verification, but may 
be the reality of flavor in the landscape.  Although the distributions are 
broad, the observed {\em patterns} among flavor parameters are present.  
For example, although the distributions of quark and charged 
lepton masses have considerable overlap within any given family, 
correlations among these masses are such that they are typically 
hierarchically separated.  Furthermore, quark mixing angles are typically 
small, and $\theta^{\rm CKM}_{13}$ tends to be smaller than the others.  It 
should be emphasized that the mass hierarchies and small mixing angles of 
the quark sector exhibit both the pairing structure and generation structure 
of the Standard Model, whereas neither of these is input directly into the 
Gaussian landscape.  

Note that the distribution of the $d$ quark Yukawa coupling (and that of the 
electron) does not range over such small values as that of the $u$ quark.
This reflects a reduced hierarchy, which is a consequence of the broader 
widths assigned to the down-type quarks (lepton doublets).  If down-type
quarks were assigned width $d$, their distributions would match those of 
up-type quarks.  As they are, the distributions do not provide a bad fit to 
observation.  However, if we consider an extension of the Gaussian landscape 
to include low-energy supersymmetry breaking, then $\tan\beta\sim \mO(10)$ 
would seem to provide an improved statistical fit.  Further investigation of 
this possibility is beyond the scope of this work.  See also 
footnote~\ref{fn4}.  

The mechanism that generates small mixing angles in the quark sector---which
stems from the up-type and down-type Yukawa overlaps sharing the localized 
wavefunctions of the quark doublets and Higgs boson---fails in the lepton 
sector, due to the broad widths of the lepton doublet wavefunctions.  
Ref.~\cite{HSW2} found that broad-width lepton doublets to be necessary but 
insufficient: without large CP-violating phases $r_a$ the relevant terms 
contributing to the PMNS matrix canceled.  We find, however, that large 
CP-violating phases are unnecessary if the kinetic terms $K^a_{ij}$ are 
determined by overlap integration, as in Eq.~(\ref{KI}).  The distributions 
of lepton mixing angles are broad, even with $r=0$, so long as 
$d_\ell/L \gtrsim 0.3$.

Ref.~\cite{HSW2} found low-energy neutrinos to exhibit enormous mass 
hierarchy when $N_n=3$.  This hierarchy ultimately stems from rank reduction 
in the mass matrix $C^\nu$, which has the effect of ``adding'' the 
hierarchies among the eigenvalues of $\lambda^n$ to those of $\lambda^M$.  
Qualitatively, the low-energy neutrino mass hierarchy is
related to the hierarchy among the three largest values of 
$\lambda_{n_i}^2/\lambda_{M_i}$, where $\lambda_{n_i}$ and $\lambda_{M_i}$ 
denote respectively the eigenvalues of $\lambda^n$ and $\lambda^M$.  When 
$N_n=3$, the hierarchy among these terms is quite large.  However, a 
consequence of choosing a large number of right-handed neutrinos, $N_n\gg3$, 
is their mass eigenvalues become densely packed.  This is not hard to 
understand:  the overall range of hierarchy is proportional to $(L/d_n)^2$, 
and as more masses are inserted in this range their typical separation 
decreases.  In turn, the three largest values of 
$\lambda_{n_i}^2/\lambda_{M_i}$ exhibit reduced hierarchy as $N_n$ is 
increased.  Indeed, we see in Fig.~\ref{fig:1} that the hierarchy among 
low-energy neutrino masses is not much larger than that among the quarks 
and charged leptons.  

Furthermore, it seems correlations tend to prevent the low-energy neutrino 
mass hierarchy from becoming too large in any given vacuum.  As an example
of this, in Fig.~\ref{fig:1} we display the distribution of (the logarithm
of) $m^\nu_2/m^\nu_3$, the ratio of the middle to the most massive 
low-energy neutrino.  We see that typically $m^\nu_3$ is not much more than 
10--100 times larger than $m^\nu_2$.  Note that, in the limit of 
hierarchical low-energy neutrino masses, 
\bea
\sqrt{\Delta m^2_\odot/\Delta m^2_{\rm atm}} 
\simeq m^\nu_2/m^\nu_3 \,.
\eea  
Hence in Fig.~\ref{fig:1} we provide the observational constraint on the 
ratio of $\Delta m^2$ of solar to atmospheric neutrino oscillations.  The 
Gaussian landscape provides a reasonable fit to this observation.  

Fig.~\ref{fig:1} refers to a specific choice of right-handed neutrino 
wavefunction widths $d_n$, as well as that of the symmetry-breaking field(s) 
$d_\phi$.  However, we have performed numerical simulations like that above 
but with $d_n/L=0.3$ or $d_\phi/L=0.3$, and have found that when $D\geq 2$ 
and $N_n\gg3$ the distributions of flavor observables are qualitatively 
unchanged from those in Fig.~\ref{fig:1}.  It should be noted that choosing 
$N_n=3$ increases the sensitivity of flavor observables to the distribution 
of right-handed neutrino masses, so that $m^\nu_2/m^\nu_3$ becomes sensitive
to $d_n$.  A larger low-energy neutrino mass hierarchy becomes more likely
in this case, with the best fit to observation coming from choosing 
narrow $d_n$.    

\begin{figure*}[t!]
\begin{center}
\begin{tabular}{c c c c} 
\includegraphics[width=0.21\linewidth]{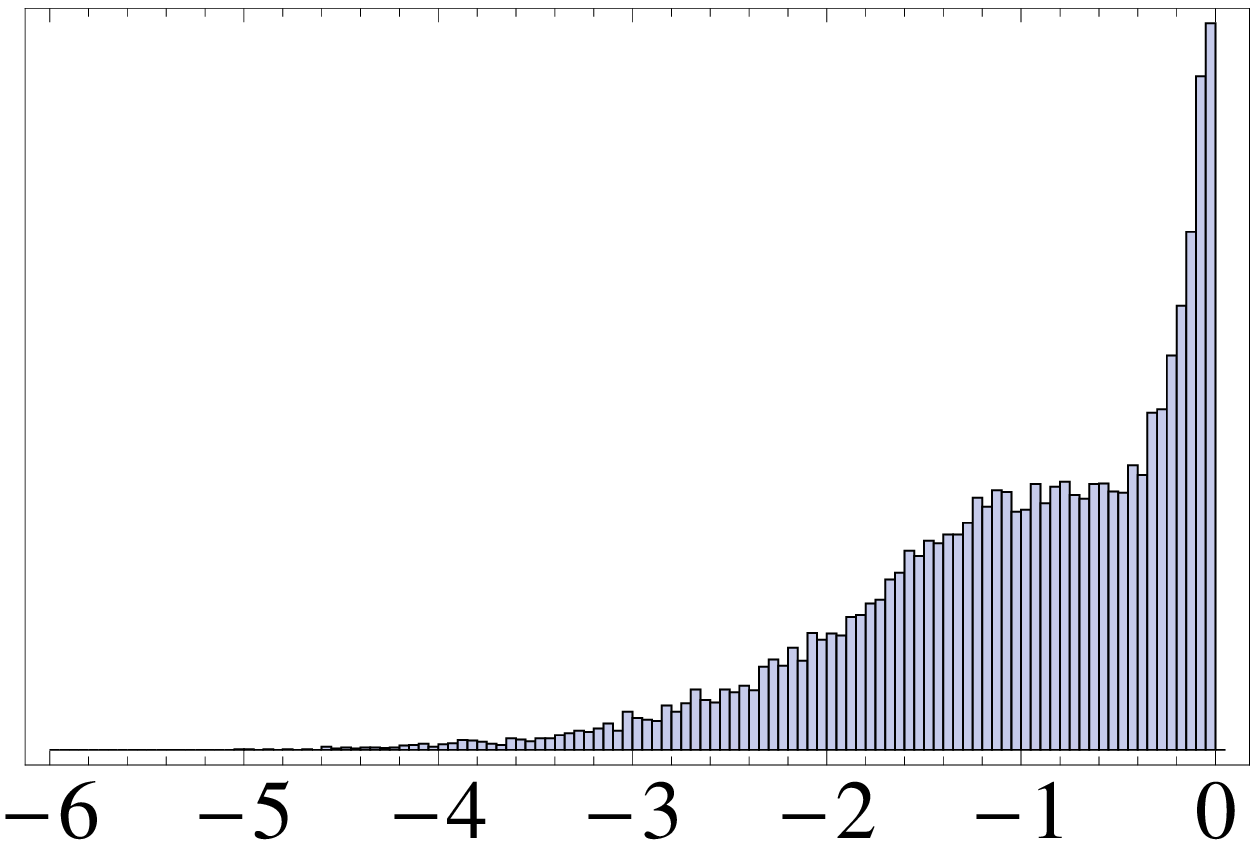} & 
\includegraphics[width=0.21\linewidth]{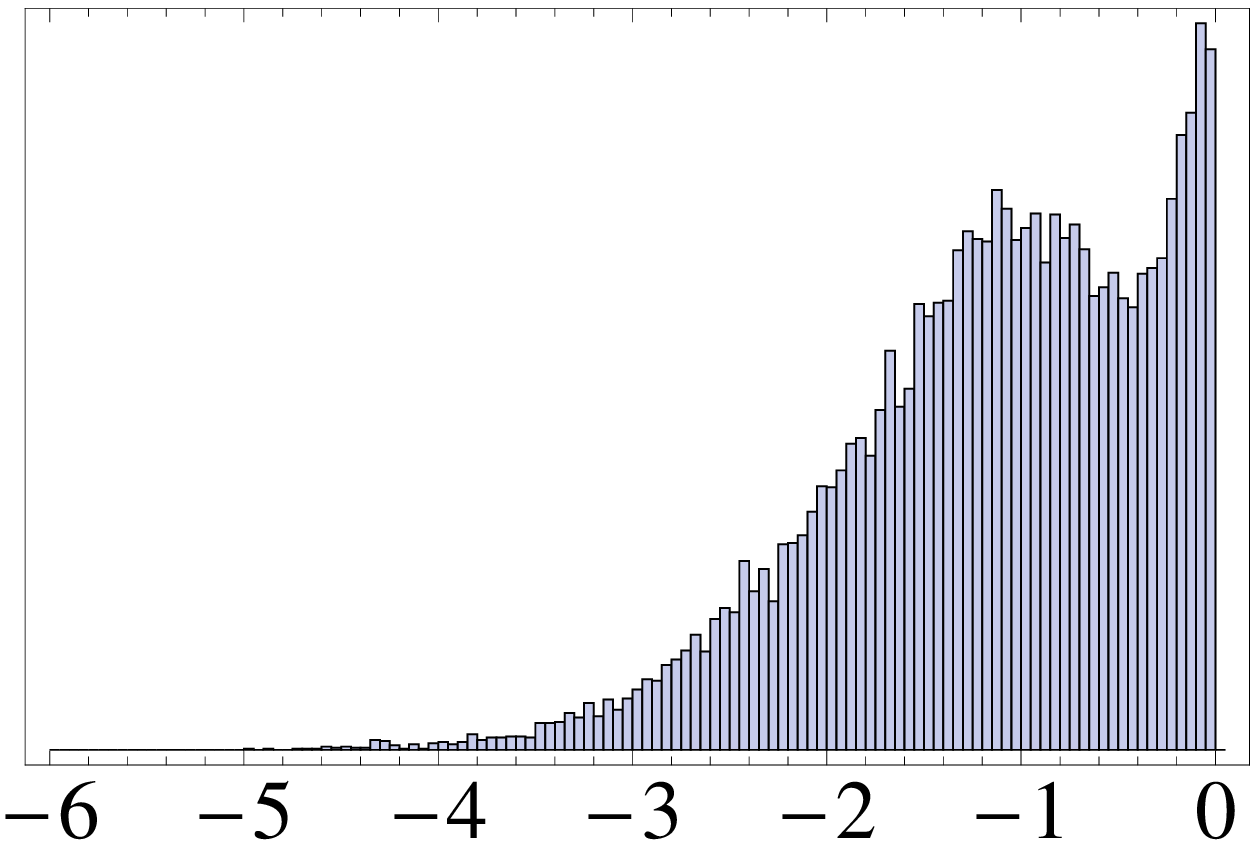} & 
\includegraphics[width=0.21\linewidth]{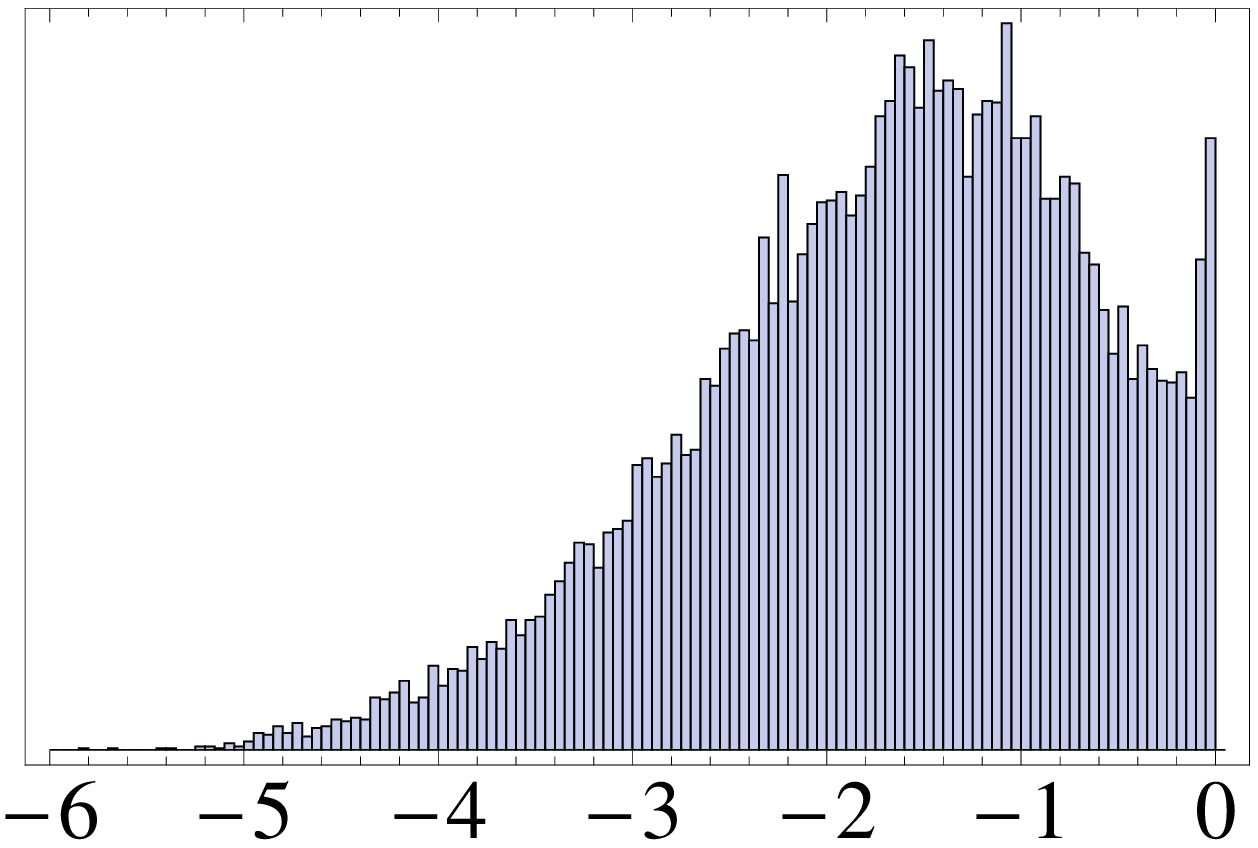} &
\includegraphics[width=0.21\linewidth]{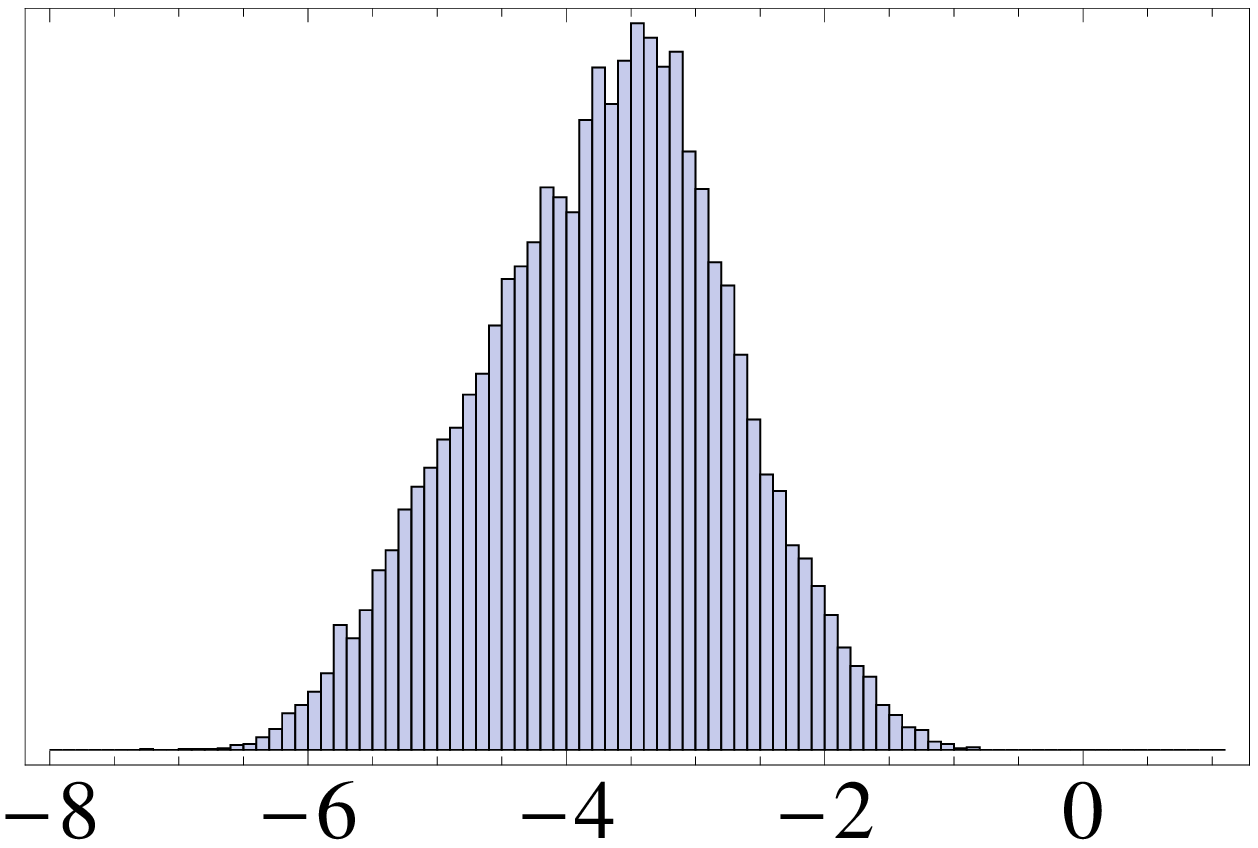} \\
$\log(2\theta^{\rm CKM}_{12}/\pi)$ [$-0.8$] &
$\log(2\theta^{\rm CKM}_{23}/\pi)$ [$-1.6$] &
$\log\sin(\theta^{\rm CKM}_{13})$ [$-2.4$] &
$\log\lambda_{d,e}$  [$-5.2,-5.5$] \\
\includegraphics[width=0.21\linewidth]{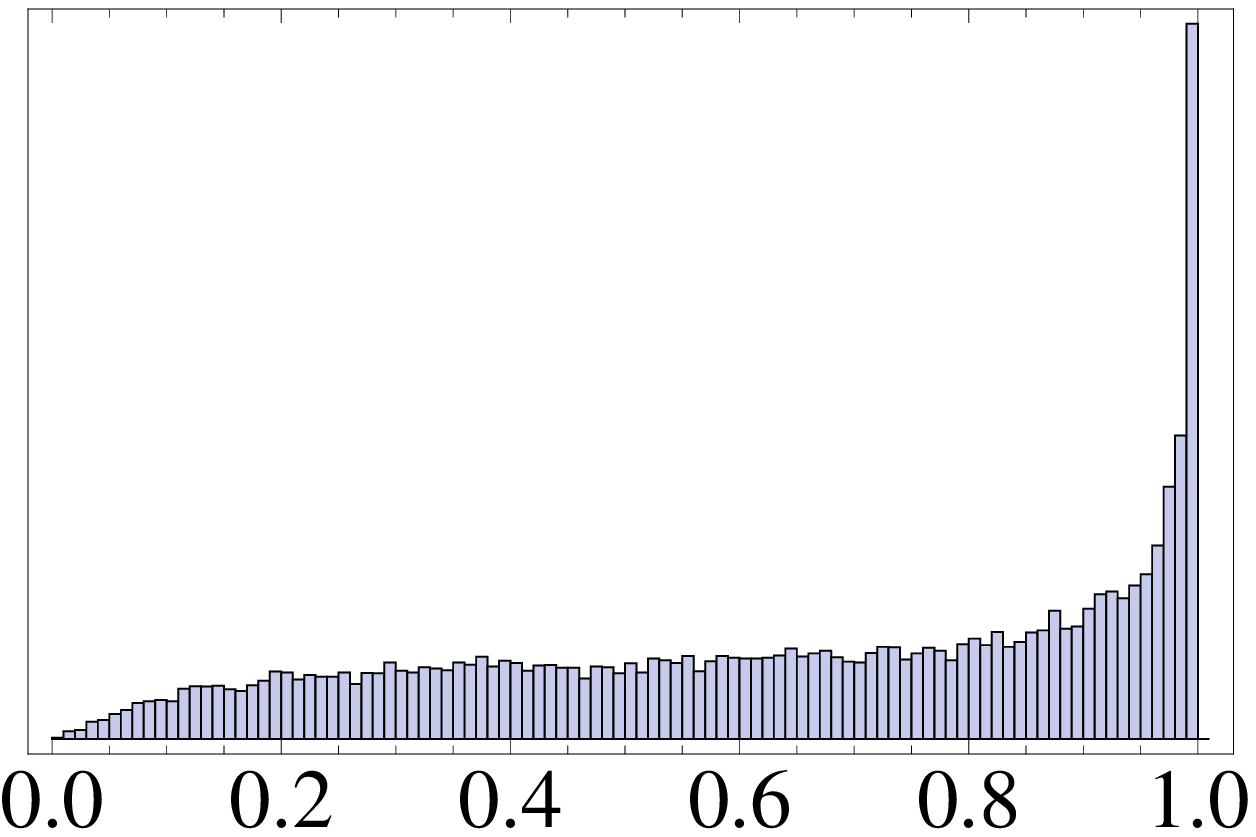} &
\includegraphics[width=0.21\linewidth]{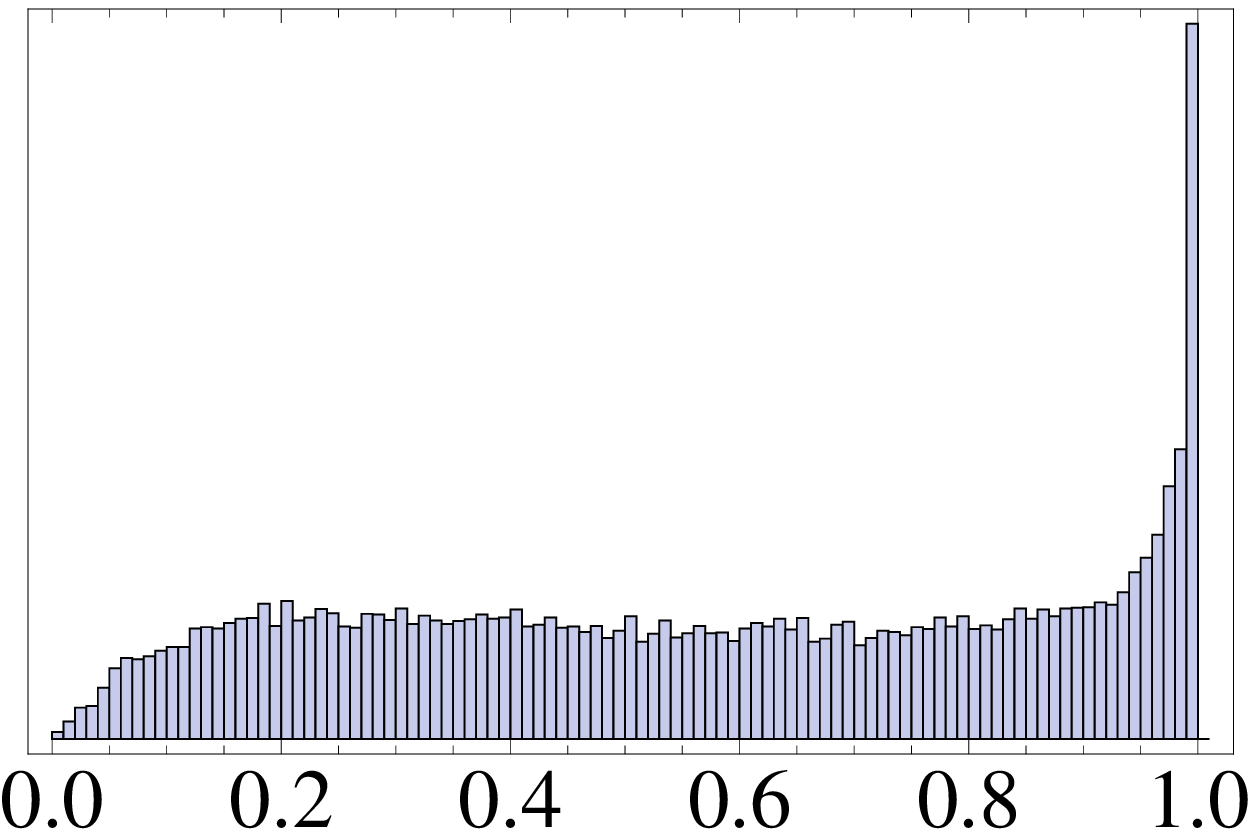} &
\includegraphics[width=0.21\linewidth]{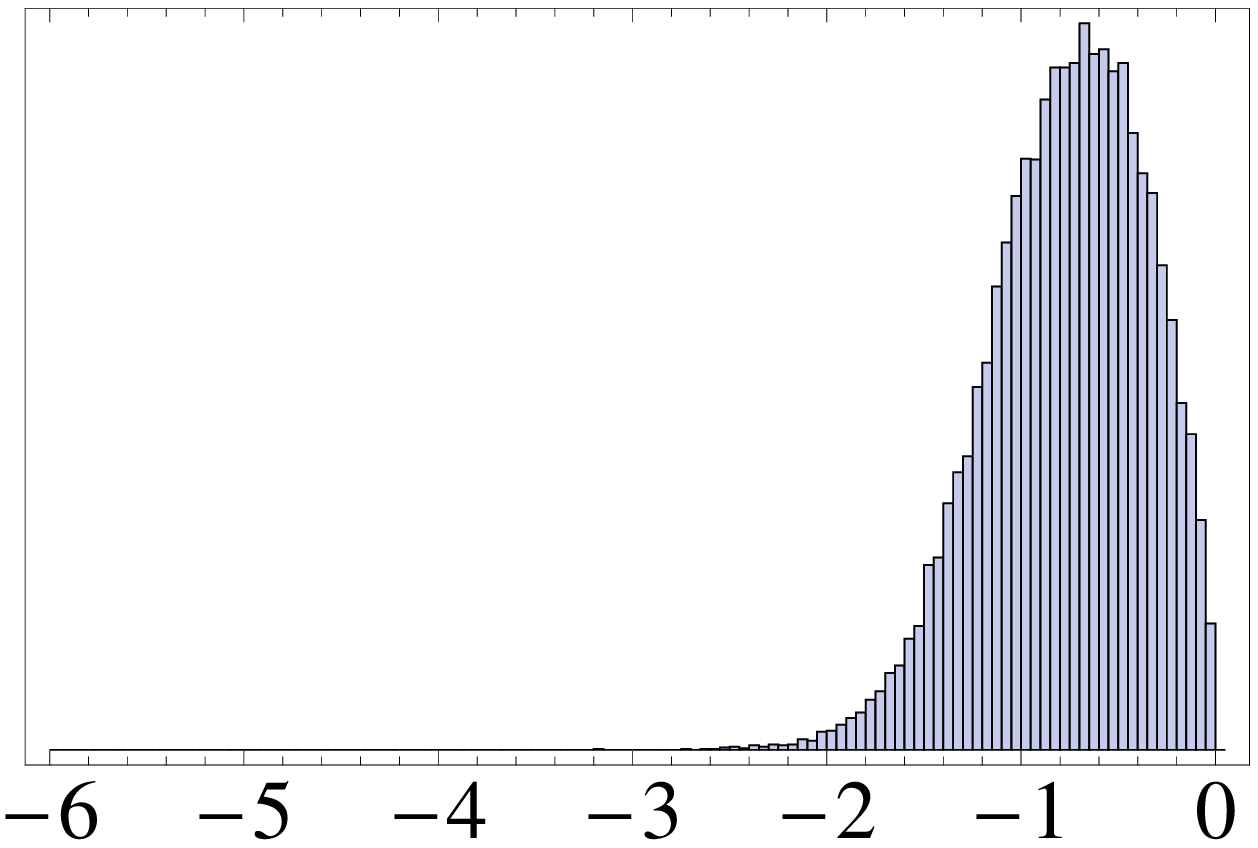} &
\includegraphics[width=0.21\linewidth]{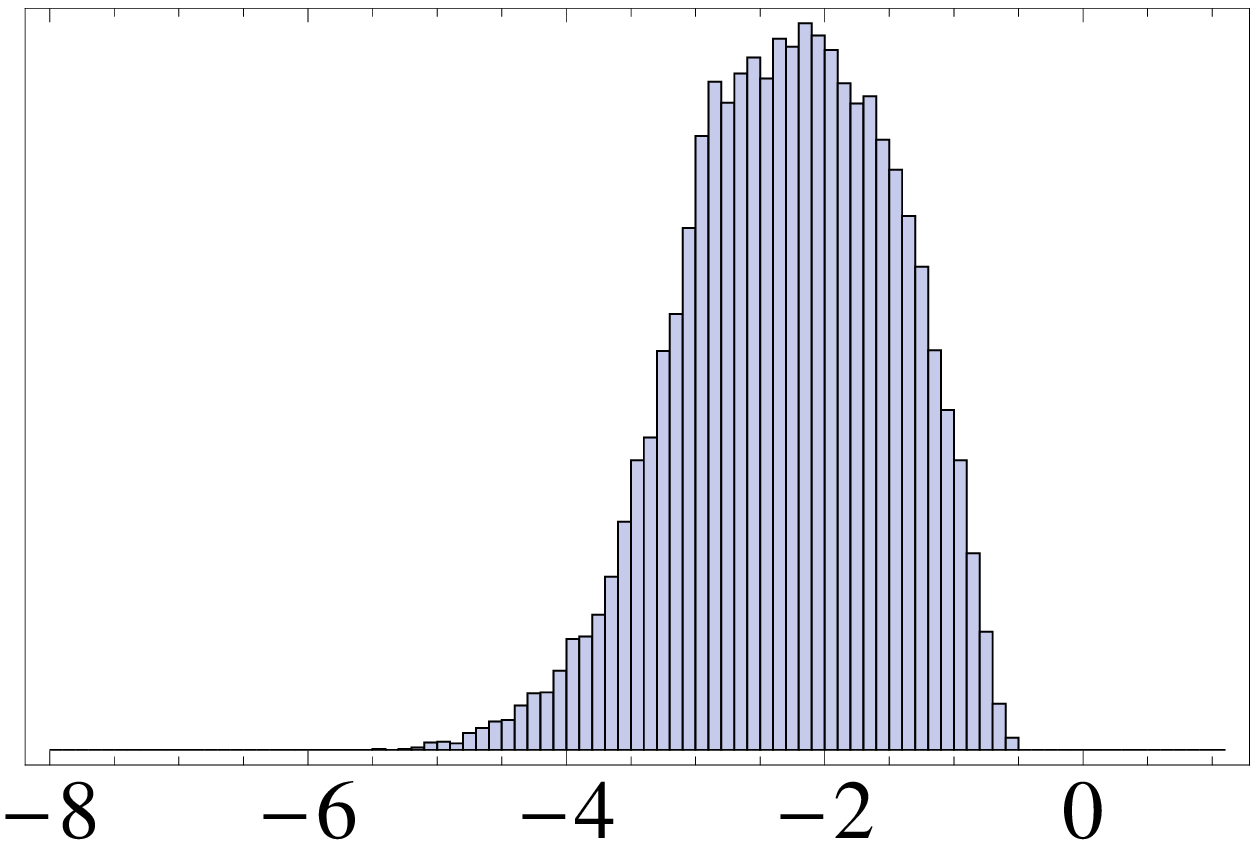} \\
$\sin(2\theta^{\rm PMNS}_{12})$ [0.91--0.93] &
$\sin(2\theta^{\rm PMNS}_{23})$ [$>0.96$] &
$\log\sin(\theta^{\rm PMNS}_{13})$ [$<-0.74$] &
$\log\lambda_{s,\mu}$ [$-3.9,-3.2$] \\
\includegraphics[width=0.21\linewidth]{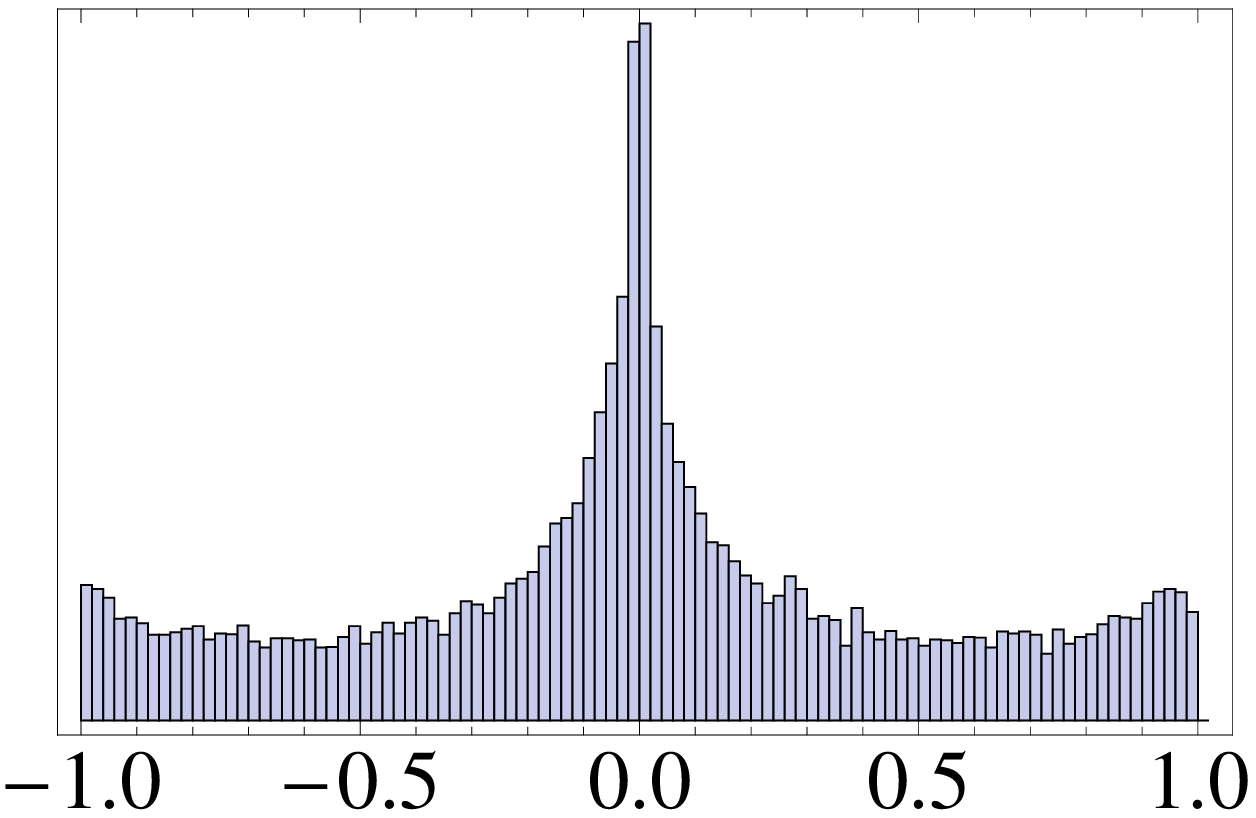} & 
\includegraphics[width=0.21\linewidth]{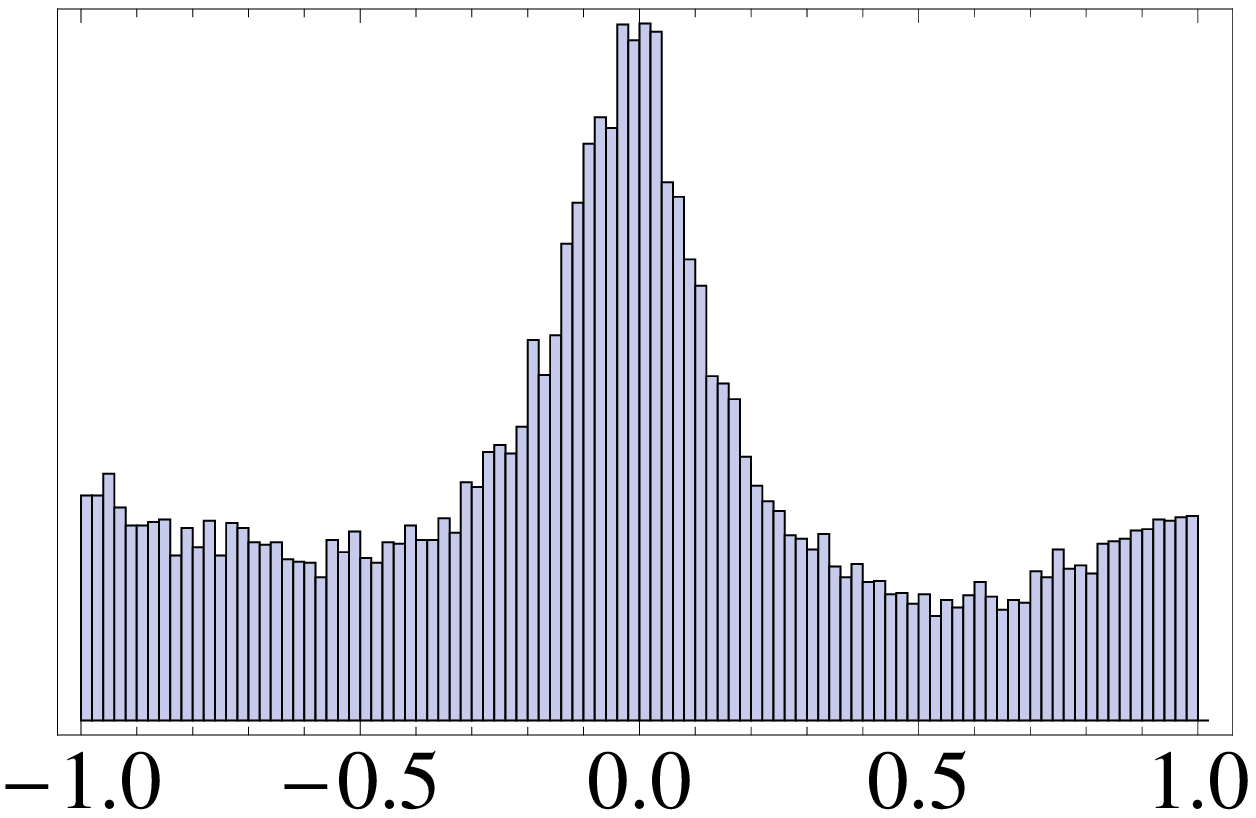} &
\includegraphics[width=0.21\linewidth]{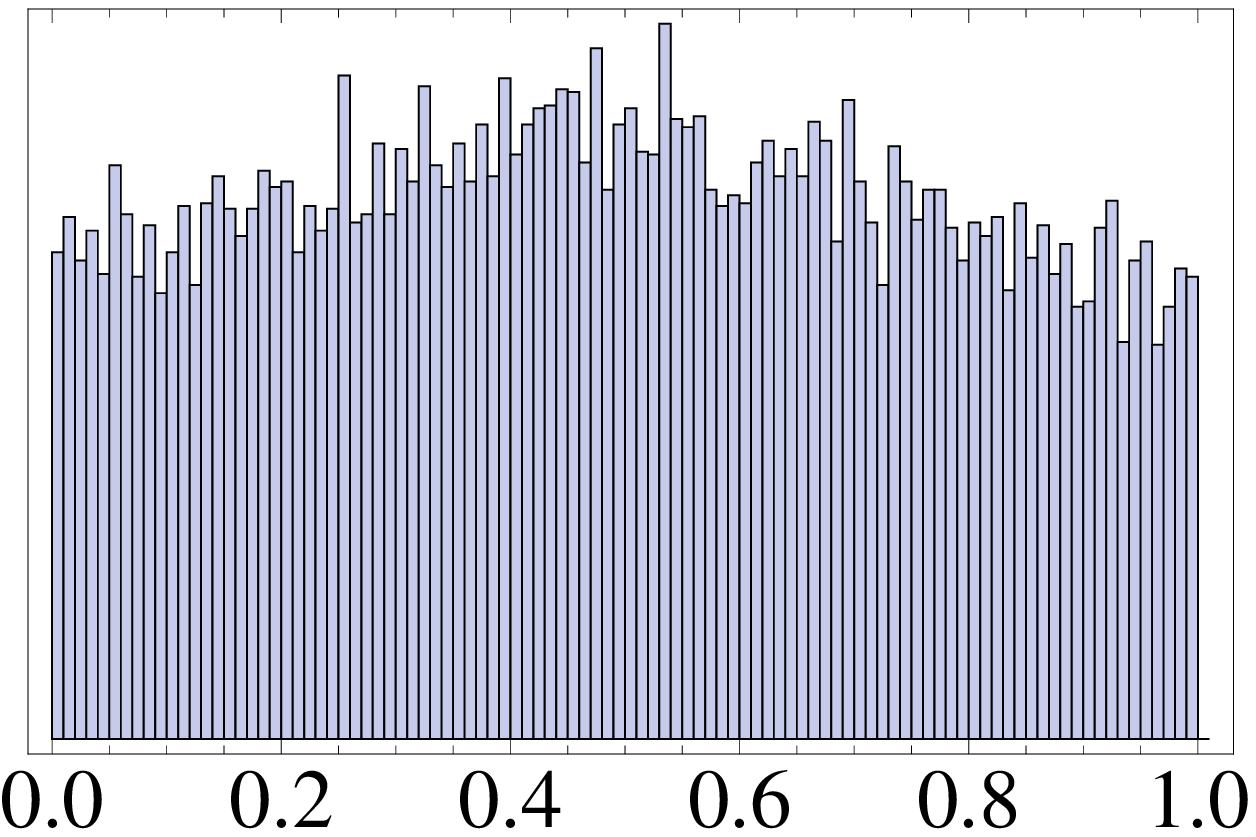} &
\includegraphics[width=0.21\linewidth]{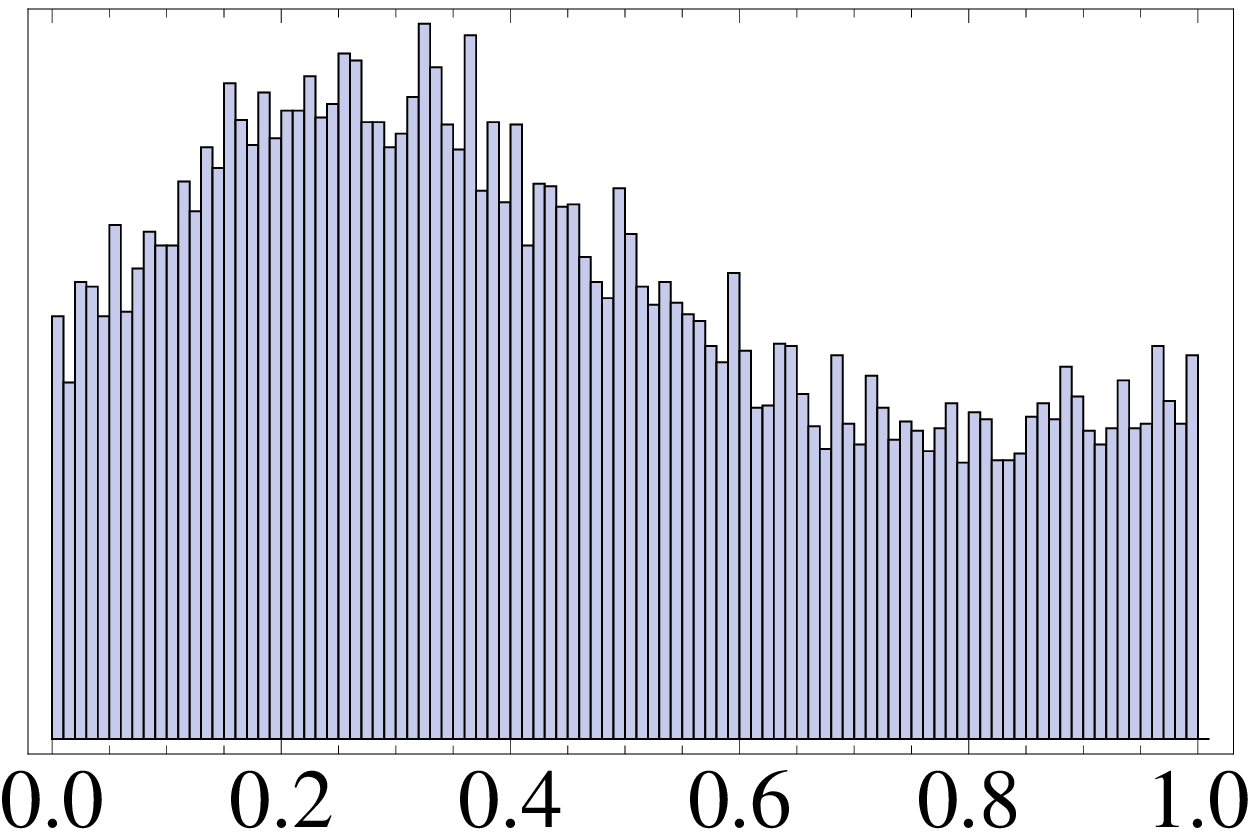} \\
$\delta^{\rm CKM}/\pi$ [0.29--0.34] & 
$\delta^{\rm PMNS}/\pi$ & 
$\alpha_1/\pi$ & 
$\alpha_2/\pi$ \\
\end{tabular}
\caption{\label{fig:2}  Distributions of flavor observables for the 
Gaussian landscape on $T^2$.  Notation and parameters are as in 
Fig.~\ref{fig:1}, except $r=0.3$.  Distributions of flavor observables that
are not displayed above are essentially unchanged from when $r=0$.}
\end{center}
\end{figure*}

Let us now discuss CP violation in the Gaussian landscape.  As has already
been remarked, it is unclear how CP violation should be introduced into the 
Gaussian landscape in order to best reflect the phenomenology of string
compactification.  We have explored the two most obvious choices---using a 
$\Gamma$-matrix phase function $p(y)$ or wavefunction phases $r_a$---and,
remarkably, have found that both leave the distributions of 
CP-conserving observables qualitatively intact.  Furthermore, at least for
the simple cases we have considered, both give similar predictions for 
the distributions of CP-violating phases.  

For illustration, in Fig.~\ref{fig:2} we display results of a numerical 
simulation exactly like that behind Fig.~\ref{fig:1}, but using $r_a=r=0.3$ 
for all $a$ (we study uniform $r_a=r$ merely for simplicity).  Some 
CP-conserving flavor observables are not displayed in Fig.~\ref{fig:2} 
because we could not detect any changes in their distributions relative to 
Fig.~\ref{fig:1}.  Those observables for which we could detect changes are 
displayed in the top two rows of Fig.~\ref{fig:2}.  It is seen that these 
changes are not very significant, considering that our universe provides 
only one data point to sample each distribution.  The four CP-violating 
phases are displayed in the bottom row.  Although the distribution of 
$\delta^{\rm CKM}$ is peaked at zero, the value we observe does not appear 
atypical.  Both increasing and decreasing $r$ from $r=0.3$ tends to sharpen 
the peak at $\delta^{\rm CKM}=0$, the effect being rather mild for increasing 
$r$.  (Increasing $r$ also tends to broaden the distributions of the quark 
mixing angles and of the mass eigenvalues $\lambda_{d_i,e_i}$, while at the 
same time pushing their medians to smaller values.)  

We have also performed a numerical simulation using $r=0$, 
$p(y)=e^{2\pi i(y_1+y_2)/L}$ (for clarity we make explicit the two 
coordinates $y_1$ and $y_2$ of the internal manifold $T^2$).  The resulting
distributions are almost indistinguishable from those obtained by choosing 
$r=0.3$, $p(y)=1$ (Fig.~\ref{fig:2}), except for the distributions of 
$\alpha_1$ and $\alpha_2$, which are completely flat, and those of some 
CP-conserving observables,  which more closely resemble the distributions 
in Fig.~\ref{fig:1}.


\section{The AFS Approximation }
\label{sec:AFSapprox}

Although computations involving overlap integrations, 
diagonalizing mass matrices, etc., are most conveniently carried out 
numerically, it is possible to understand analytically the qualitative 
features of the distributions of various flavor observables in Gaussian 
landscapes.  Such an understanding develops intuition for how these 
distributions depend on the values of parameters that were taken as given
in the numerical simulations of Section~\ref{sec:numerical}, and thus 
allows us to anticipate the results of other choices of parameters without 
having to repeat the numerical simulation.  Furthermore, as was seen in 
Ref.~\cite{HSW2}, such an analysis can reveal how the distributions of 
flavor observables are related to various features of the geometry
of compactification.  Thus, we here extend the analysis of Ref.~\cite{HSW2}
by taking account of the kinetic mixing in the $K^a_{ij}$ matrices.

In Ref.~\cite{HSW2}, the present authors found that the distributions 
of Yukawa couplings in Gaussian landscapes (using the basis of zero modes 
explored there) are almost the same as those predicted by models with 
approximate abelian flavor symmetries (AFS).  In the AFS approach to 
flavor~\cite{FN}, one assumes each of the Standard--Model fermions has its 
own U(1) charge.  If the $i$th anti-up quark has a U(1) charge 
$a_i \geq 0$, the $j$th left-handed quark doublet has a charge 
$b_j \geq 0$, and the Higgs boson is neutral under the U(1) symmetry, then 
their Yukawa coupling is  
\bea
\lambda^u_{ij} = g_{ij} \epsilon^{a_i} \epsilon^{b_j} \,,
\label{AFSdem}
\eea
where the small parameter $\epsilon \ll 1$ comes from a small breaking of 
the U(1) symmetry, and the $g_{ij}$ are coefficients of order unity, 
presumably determined by a higher-energy theory.  Ref.~\cite{HSW2}
showed that Yukawa couplings have the same structure in Gaussian landscapes, 
when the small suppression factors $\epsilon^{a_i}$ and $\epsilon^{b_j}$ are 
interpreted properly, even though there is no approximately preserved U(1) 
flavor symmetry.  This allows for a simple analytic understanding of
quark and lepton masses and mixings.  Below we show that a similar 
analytic understanding of the predictions of Gaussian landscapes is 
possible when including non-trivial kinetic mixing $K^a_{ij}$ and 
increasing the number of right-handed neutrinos.

To proceed, we introduce a series of simplifying approximations~\cite{HSW2}.  
First, we ignore the fact that $B$ is a compact manifold and instead treat 
it like Euclidian space $\R^D$ when calculating overlap integrals.  This 
approximation becomes more accurate in the limit where wavefunctions are 
peaked nearer to each other---i.e.~when the Yukawa couplings become 
large---but is surprisingly effective at describing the qualitative shapes 
of distributions of Yukawa matrix elements, as was confirmed by a number 
of numerical simulations in Ref.~\cite{HSW2}.\footnote{It is interesting 
that curvature of $B$ may introduce a statistical bias.  We do not pursue
this possibility here.}  Furthermore, for simplicity 
we set $p(y)=1$ and $r_a=0$ (recall that the CP-conserving observables are 
approximately independent of the $r_a$ when $r_a\lesssim0.3$).  

With these approximations, the relevant overlap integrals have simple 
analytic expressions.  For example, the up-type Yukawa matrix is given by
\bea
\lambda^{u}_{ij} \sim \exp\!\left[
-\frac{y_{u_i}^2+y_{q_j}^2-y^{u_i}y^{q_j}}{3d^2}\right] ,
\label{AFS}
\eea
where we have used translational invariance over the extra-dimensional 
space $\R^D$ (this is true in $T^D$ as well) to set $y^h=0$.  Note that for 
$D>1$, the $y^{a_i}$ are vectors on the space of extra dimensions, and terms 
like $y^{a_i}y^{b_j}$ correspond to scalar products of vectors.  The last 
term in the exponent is ``statistically neutral;'' that is, it is linear in 
the random peak positions $y^{u_i}$ and $y^{q_j}$, and therefore can be 
positive or negative, with a contribution that largely cancels in the 
statistical distribution.  

The up-type Yukawa matrix has the AFS structure, 
\bea
\lambda^u_{ij} \sim g_{ij} \epsilon^u_i \epsilon^q_j,
\label{eq:AFS-u}
\eea
where the small factors $\epsilon^u_i$ and $\epsilon^q_j$ correspond 
to the first two terms in the exponent of $\lambda^u_{ij}$ in 
Eq.~(\ref{AFS}), while the $g_{ij}$ comes from the third term.  Again, 
the $g_{ij}$ are random numbers that may be larger or smaller than unity, 
but are statistically neutral.  We can always relabel the generation 
index $i$ of the three independent anti-up-type quarks $\bar{u}_i$, 
so that $\epsilon^u_1 < \epsilon^u_2 < \epsilon^u_3$.  Similarly, 
the generation index for the three quark doublets can be relabeled, so 
that $\epsilon^q_1 < \epsilon^q_2 < \epsilon^q_3$.  
Note that in approximate AFS models, $\epsilon$ is a fixed-value
symmetry-breaking parameter, and the U(1) charge assignments $a_i$ and 
$b_j$ are picked ``by hand,'' whereas in Gaussian landscapes the 
``AFS factors'' $\epsilon^u_1 < \epsilon^u_2 < \epsilon^u_3$ and 
$\epsilon^q_1 < \epsilon^q_2 < \epsilon^q_3$ follow statistical 
distributions.  The quantity $(L/d)^2$ sets the order-of-magnitude of 
hierarchy in Gaussian landscapes, $\lambda \sim e^{-(L/d)^2}$, playing 
a role analogous to the symmetry-breaking parameter $\epsilon$ in  
approximate AFS models.

For relevance to the results of Section~\ref{sec:numerical}, we assume 
the down-type quark wavefunctions have a relatively broad width 
$d_{\bf \bar{5}}\sim\mO(L)$, while all other fields have the same (narrow) 
width $d\ll L$.  The down-type Yukawa matrix then has the structure
\bea
\lambda^d_{ij} \sim \varphi^d_i|_{y=y^{q_j}\!/2}\,
\left(\epsilon^q_j\right)^{\frac{3}{4}} .
\label{eq:AFS-d}
\eea
The broad-width wavefunctions $\varphi^d_i$ are simply evaluated at the
peak of the overlap, which occurs between the localization of the Higgs 
boson $y=y^h=0$ and that of the quark-doublet wavefunctions $y=y^{q_j}$, 
and can be seen to introduce a set of random numbers of order the 
down-type wavefunction normalization.  Thus, the down-type Yukawa matrix 
$\lambda^d_{ij}$ also follows the structure of approximate AFS models.  

So far, our discussion has mirrored that of Refs.~\cite{HSW1,HSW2}.  
However, that work ignored the effect of kinetic mixing.  Applying the
approximations described above, the kinetic mixing of quarks is given by
\bea
 K^a_{ij} \sim 
 \exp\!\left[-\frac{1}{4d^2}\left(y^{a_i}-y^{a_j}\right)^2\right] \,,  
\eea
for $a=u,d,q$.  Such a kinetic matrix can be put in canonical form by
rotating quark flavors using a ``triangular'' matrix, for example the 
quark doublets are made canonical (with an appropriate normalization 
of $\varphi^q_i$) by       
\bea
\left(\begin{array}{c}
q_1 \\ q_2 \\ q_3
\end{array}\right) \sim 
\left(\begin{array}{ccc}
1\,\, & -K^q_{12} & -K^q_{13} \\
0\,\, & 1 & -K^q_{23} \\
0\,\, & 0 & 1
\end{array}\right) 
\left(\begin{array}{c}
q'_1 \\ q'_2 \\ q'_3
\end{array}\right) \,.
\label{TN}
\eea
In the above, we have kept only leading order terms; note $K^q_{ij}$ is
of order $(\epsilon^q_i \epsilon^q_j)^{\frac{3}{4}}$ for $i\ne j$ and 
thus the off-diagonal terms are small compared to unity.  An analogous 
expression applies to the up-type quarks.  The down-type kinetic 
matrices do not necessarily have small off-diagonal terms, yet they 
can be put in canonical form via a matrix of such ``triangular'' form.

Note that these rotations do not affect the AFS structure of the quark
Yukawa couplings.  The off-diagonal terms in the ``triangular'' rotation
matrices for up-type quarks and quark doublets are negligible in this
qualitative analysis, and so $\lambda^u$ retains the structure of 
Eq.~(\ref{eq:AFS-u}).  Meanwhile, the ``triangular'' form of the 
down-type quark rotation essentially replaces the random terms 
$\varphi^d_i|_{y=y^{q_j}/2}$ with linear combinations of such 
terms---this does not change the structure of $\lambda^d$.  Thus, the
quark sector of the Gaussian landscape retains its AFS structure.  It is 
straightforward to see that the Yukawa eigenvalues are roughly given by 
$\epsilon^u_i\epsilon^q_i$ ($i=1,2,3$), and $(\epsilon^q_j)^{\frac{3}{4}}$ 
($j=1,2,3$), in the up-type and down-type sectors, respectively.

As has been mentioned, the AFS factors $\epsilon^u_i$ and $\epsilon^q_j$
are described by correlated statistical distributions reflecting the 
random scanning of Gaussian wavefunction peak positions.  Their 
statistical distributions encapsulate the hierarchical suppression that
must be inserted ``by hand'' using U(1) flavor charges in AFS models.   
Ref.~\cite{HSW2} explains how the geometry of $B$ affects the 
distributions of the AFS factors.  That analysis carries over directly
to the Gaussian landscapes considered here, so for brevity we do not 
repeat it.  Instead, we simply note that distributions of flavor 
observables are qualitatively independent the geometry of the internal
manifold.   

Let us now turn to quark-sector electroweak mixing.  Keeping in mind 
that kinetic mixing does not change the structures of $\lambda^u$ and 
$\lambda^d$ in Eqs.~(\ref{eq:AFS-u}) and~(\ref{eq:AFS-d}), the CKM 
mixing angles are given by the difference between the diagonalization 
angles $\theta_{ij} \sim \epsilon^q_i/\epsilon^q_j$ ($i<j$) for the 
up-type Yukawa matrix and 
$\theta_{ij} \sim (\epsilon^q_i/\epsilon^q_j)^{\frac{3}{4}}$ ($i<j$) 
for the down-type Yukawa matrix.  Qualitatively, the difference is 
dominated by the larger mixing angle, and hence   
\bea
\theta^{\rm CKM}_{ij} 
\sim (\epsilon^q_i/\epsilon^q_j)^{\frac{3}{4}}
\sim \lambda_{d_i}/\lambda_{d_j}
\sim (\lambda_{u_i}/\lambda_{u_j})^{\frac{3}{8}} \,.
\eea
The last two expressions relate the typical size of CKM mixing angles
to typical ratios of quark mass eigenvalues, due to all of these 
observables being qualitatively determined by the same AFS factors.    

The lepton sector can be analyzed similarly.  Now the lepton 
doublet has broad width $d_{\bf\bar{5}}$, while the other zero modes
have narrow wavefunctions.  Thus, the charged-lepton and neutrino
Yukawa matrices have the structure  
\bea
\lambda^e_{ij} & \sim & (\epsilon^e_i)^{\frac{3}{4}}\,
\varphi^\ell_j|_{y=y^{e_i}\!/2} \\
\lambda^n_{ij} & \sim & (\epsilon^n_i)^{\frac{3}{4}}\,
\varphi^\ell_j|_{y=y^{n_i}\!/2} \,. 
\eea
These matrices have essentially the same structure as 
$\lambda^d_{ij}$.  Likewise, kinetic mixing in the lepton sector is
either negligible (for electroweak singlets) or simply mixes the
random factors $\varphi^\ell_j$.  Hence, the mass eigenvalues in the
charged-lepton sector are qualitatively given by 
$(\epsilon^e_1)^{\frac{3}{4}}$, $(\epsilon^e_2)^{\frac{3}{4}}$, and
$(\epsilon^e_3)^{\frac{3}{4}}$, where again we relabel generation
indices so that $\epsilon^e_1 < \epsilon^e_2 < \epsilon^e_3$.  Their
distributions are the same as those of the down-type quarks.  

Because the right-handed neutrino mass matrix $\lambda^M$ involves the
overlap of two right-handed neutrinos, it is well approximated as a 
diagonal matrix,
\bea
\lambda^M_{ij} \sim \epsilon^M_i\delta_{ij} \,.
\eea
The diagonal elements $\epsilon^M_i$ may be small due to small 
right-handed neutrino overlap with the symmetry-breaking field 
$\phi$.  The seesaw mass matrix of left-handed neutrinos then 
takes the form
\bea
C_{ij} \sim \sum_{k=1}^{N_n} 
\varphi^\ell_i|_{y=y^{n_k}\!/2}\,\, 
\varphi^\ell_j|_{y=y^{n_k}\!/2}\, 
(\epsilon^n_k)^{\frac{3}{2}}\!/\epsilon^M_k \,.
\eea
The distribution of AFS factors $\epsilon^n_k\sim e^{-(y^{n_k})^2\!/3d^2}$ 
covers a logarithmic range proportional to $(L/d)^2$.  If, as we have 
assumed, the Majorana mass matrix $\lambda^M_{ij}$ is determined by overlap 
integration of right-handed neutrinos with a Gaussian wavefunction for 
$\phi$, then the diagonal elements $\epsilon^M_i$ are also distributed over
a logarithmic range proportional to $(L/d)^2$.  The seesaw mass matrix 
$C_{ij}$ involves summation over all right-handed neutrinos, but its 
three eigenvalues are essentially determined by the contributions with the
three largest values of 
$c_k\equiv (\epsilon^n_k)^{\frac{3}{2}}\!/\epsilon^M_k$.  Although 
collectively the $N_n$ individual $c_k$ range over a logarithmic scale 
proportional to $(L/d)^2$, the {\em three largest} values strongly
depend on $N_n$.  As the number of right-handed neutrinos $N_n$ is 
increased, typically the three largest values of $c_k$ will become 
bunched near the largest possible value of $c_k$.  In turn, the three
eigenvalues of $C_{ij}$ experience reduced hierarchy as $N_n$ is 
increased.  This is in good agreement with the measured ratio of 
$\Delta m^2$ coming from solar and atmospheric neutrino oscillations.
 
Finally, we consider the distributions of the PMNS mixing angles.  
Because the different elements of a given row of $\lambda^e$ and
$C$ are comparable to each other (due to broad-width lepton doublet
wavefunctions), large-angle left-side rotations are necessary to 
diagonalize these matrices.  Naively, these large-angle rotations 
would correspond to large PMNS mixing angles.\footnote{Note that this
argument does not apply to the down-type quark mass matrix, with regard
to the CKM mixing angles, because $\lambda^d$ has hierarchical rows.  
The different elements of a given {\em column} of $\lambda^d$ are 
comparable, and this translates to large-angle right-side 
diagonalization rotations, but such rotations do not enter into the 
CKM matrix.}  However, the present authors concluded in Ref.~\cite{HSW2} 
(where kinetic mixing was ignored) that broad-width lepton doublet 
wavefunctions are insufficient to generate large PMNS mixing angles.  
The reason is, the broad-width wavefunctions $\varphi^\ell_j(y)$ are 
not very different for different zero modes (different $j$).  As a 
result, the diagonalization angles for the left-handed charged leptons 
and those for the left-handed neutrinos are highly constrained, and 
an accidental cancellation occurs in the PMNS matrix.  Thus, it was 
concluded that some other ingredient(s)---such as large CP-violating 
phases---must be included in order for large lepton mixing to be 
typical in Gaussian landscapes. 

However, Ref.~\cite{HSW2} ignored kinetic mixing.  As has been remarked,
rotating leptons to a canonical basis of zero modes does not change the
qualitative structure of the Yukawa matrices $\lambda^e$ and $\lambda^n$,
nor the mass matrices $\lambda^M$ and $C$.  However, in such a basis the 
three independent zero modes of each family are orthogonal to one
another, which is quite unlike the situation imagined in Ref.~\cite{HSW2},
where these zero modes were rather homogeneous and similar.  The effect
of this is to prevent the accidental cancellation that was observed in 
Ref.~\cite{HSW2}.  Thus, PMNS mixing angles obtain their ``naive'' 
values, comparable to ratios of different elements in a given row of
$\lambda^e$ or $\lambda^n$, which are typically order unity.  Numerical
simulations in Section~\ref{sec:numerical} confirm this idea.

\section{Conclusions}
\label{sec:conclusions}

The idea that characteristics of our universe are selected from a 
landscape of vacua has proved quite provocative.  The major criticisms
of this hypothesis can be classified into three groups:  $i$) the 
landscape combined with eternal inflation implies a multiverse with 
diverging spacetime volume, expectations for observables depend on how
this diverging volume is regulated, yet it is unclear how this should be
done; $ii$) expectations for observables depend on anthropic selection
effects, however it is unclear how to define our reference class, and 
at the same time the calculation of anthropic effects tends to be very 
complicated; and $iii$) our universe provides only one data point to 
compare to the predicted distribution of each scanning parameter, making
landscape models seem difficult to falsify.    

On the other hand, the flavor physics of a landscape tends to avoid
each of these criticisms.  Although the cosmology of eternal inflation
and the spacetime measure of the multiverse ($i$) strongly influence the 
distributions of cosmological observables such as the observed primordial
cosmic density contrast, insofar as the landscape is large enough that 
low-energy physics is decoupled from inflationary dynamics, flavor 
physics should be independent of these issues.  Furthermore, although it 
is expected that anthropic selection ($ii$) plays an important role in 
determining some flavor parameters, such as the $u$, $d$, and $e$ masses,
most flavor parameters seem decoupled from conditions necessary to 
allow for the evolution of observers.  Finally, although we have only 
one universe in which we can perform measurements ($iii$), there are 
almost two dozen (possibly correlated) flavor observables that should be 
described by the landscape model, allowing for a more discerning 
statistical analysis.  Thus, we consider the issue of flavor to be ideal 
for applying the landscape hypothesis.

\begin{acknowledgments}
This work was supported in part by the Director, Office of Science, Office
of High Energy Physics, of the U.S. Department of Energy under Contract
No. DE-AC02-05CH11231 (LJH), by the U.S. National Science Foundation 
under grants PHY-04-57315 (LJH) and NSF 322 (MPS), and by the WPI 
Initiative, MEXT, Japan (TW).
\end{acknowledgments}

\end{document}